\documentclass[]{aastex631}

\usepackage{graphicx}
\usepackage{xcolor}
\usepackage{amsmath,amssymb,amsthm}
\usepackage{mathtools}
\usepackage{CJK}
\usepackage{booktabs}
\usepackage[flushleft]{threeparttable}
\usepackage{mathtools}
\usepackage{multirow}

\newcommand{\fiesta}{$\mathit{\phi}$ESTA}
\newcommand{\FIESTA}{$\mathit{\phi}$ESTA }
\newcommand{\FIESTAtitle}{\texorpdfstring{$\mathit{\phi}$ESTA}{Lg} }
\newcommand{\Scalpels}{SCALPELS } % Lookup how to do the funny caps thing

\definecolor{amethyst}{rgb}{0.6, 0.4, 0.8}

\shorttitle{FIESTA II}
\shortauthors{Zhao, Ford \& Tinney}
\graphicspath{{./}{figures/}}
\begin{document}
\begin{CJK*}{UTF8}{gbsn}

\title{FIESTA II. Disentangling stellar and instrumental variability from exoplanetary Doppler shifts in Fourier domain}

\correspondingauthor{J. Zhao}
\email{jzhao@psu.edu}
\author[0000-0001-5290-2952]{J. Zhao}
\affiliation{Department of Astronomy \& Astrophysics, The Pennsylvania State University, 525 Davey Lab, University Park, PA 16802, USA}
\affiliation{Exoplanetary Science at UNSW, School of Physics, UNSW Sydney, NSW 2052, Australia}

\author[0000-0001-6545-639X]{Eric B. Ford}
\affiliation{Department of Astronomy \& Astrophysics, The Pennsylvania State University, 525 Davey Lab, University Park, PA 16802, USA}
\affiliation{Center for Exoplanets \& Habitable Worlds, 525 Davey Laboratory, The Pennsylvania State University, University Park, PA 16802, USA}
\affiliation{Center for Astrostatistics, 525 Davey Laboratory, The Pennsylvania State University, University Park, PA 16802, USA}
\affiliation{Institute for Computational \& Data Sciences, 525 Davey Laboratory, The Pennsylvania State University, University Park, PA 16802, USA}
\author[0000-0002-7595-0970]{C. G. Tinney}
\affiliation{Exoplanetary Science at UNSW, School of Physics, UNSW Sydney, NSW 2052, Australia}
\affiliation{Australian Centre for Astrobiology, UNSW Sydney, NSW 2052, Australia}



\begin{abstract}

The radial velocity detection of exoplanets is challenged by stellar spectroscopic variability that can mimic the presence of planets and by instrumental instability that can further obscure the detection.
Both stellar and instrumental changes can distort the spectral line profiles and be misinterpreted as apparent RV shifts.
%
% Of the numerous processes that contribute to stellar variability, stellar activity is particularly concerning, since the stellar rotation period and the active region evolution timescale are comparable to the orbit periods of common exoplanets.  Stellar activity introduces additional inhomogeneities in the surface brightness of the stellar surface.
%

We present an improved FourIEr \textit{phase} SpecTrum Analysis (FIESTA a.k.a. \fiesta) to disentangle apparent velocity shifts due to a line deformation from a true Doppler shift.  
\FIESTA projects stellar spectrum's cross correlation function onto a truncated set of Fourier basis functions.
Using the amplitude and phase information from each Fourier mode, we can trace the line variability at different CCF width scales to robustly identify and mitigate multiple sources of RV contamination. 
For example, in our study of the 3 years of HARPS-N solar data, \FIESTA reveals the solar rotational effect, the long-term trend due to solar magnetic cycle, instrumental instability and apparent solar rotation rate changes. 
Applying a multiple linear regression model on \FIESTA metrics, we reduce the weighted root-mean-square noise from 1.89\,m/s to 0.98\,m/s. In addition, we observe a $\sim$3 days lag in the \FIESTA metrics, similar to the findings from previous studies on BIS and FWHM.

\end{abstract}



%%%%%%%%%%%%%%%%%%%%%%%%%%%%%%%%%%%%%%%%%%%%%%%%%%%%%
\section{Introduction} \label{sec:intro}
%%%%%%%%%%%%%%%%%%%%%%%%%%%%%%%%%%%%%%%%%%%%%%%%%%%%%
In radial velocity exoplanet surveys, the Doppler shifts that result from the gravitational interaction of planets and their host star are measured via high-resolution spectroscopy to detect and characterise the masses and orbits of planets.  However, the spectral signature of a Doppler shift can be mimicked by deformation of the host star's spectral line profiles. 

As the field pushes towards improved Doppler precision, developing robust and powerful approaches for characterizing line profile variations, whether they be due to intrinsic stellar variability or instrumental effects, is becoming increasingly critical.
Early planet-hunting spectrographs went to great lengths to provide a precise and stable wavelength calibration, but did not stabilise the instrument, leading to significant instrumental line spread function variations. Recently, a new generation of spectrographs (e.g., ESPRESSO \citep{Pepe2014}, NEID \citep{Schwab2019}, the EXPRES Spectrometer \citep{Petersburg2020}) have been designed to be much more intrinsically stable, in an effort to enable extremely precise radial velocity (EPRV) measurements.
These instrument specifications aim to reach 10-30~cm/s radial velocity precision, with the long-term goal of characterizing potentially Earth-like planets around Sun-like stars \citep{Fischer2016}. 
To reach this goal, these instruments incorporate multiple environmental control strategies, in an effort to stabilise the line-spread function and minimise instrument-induced variability.  Equally, or perhaps even more importantly, this improved instrumental stability makes it feasible to use line shape variations as a diagnostic to recognise and mitigate the effects of stellar variability.  

Astronomers have been exploring a variety of strategies to recognise and mitigate the effects of stellar variability, ranging from looking for correlations with traditional activity indicators such as the equivalent width of H$\alpha$ \citep{Herbst1989}, the line profile full width at half-maximum \citep[FWHM,][]{Queloz2009}, the line profile's bisector \citep{Queloz2001Bis}, and data-driven and machine learning methods  (\citealp[e.g.][]{Pearson_2018, debeurs2020identifying}).

\citet{jzhao2020} proposed \FIESTA for characterizing a signal deformations and signal shifts. In the context of radial velocity detection of exoplanets, the signal is usually the cross-correlation function (CCF) of the stellar spectrum with a template spectrum or synthetic mask. By combining information from many spectral lines, the CCF has significantly higher signal-to-noise ratio (SNR) than individual spectral lines. In this study, we validate and test \FIESTA using CCFs based on a weighted binary mask \citep{Baranne1996, Pepe2002}.

\FIESTA decomposes the CCF into the orthogonal Fourier basis functions and calculates the shift for each basis function. A pure Doppler shift results in all the Fourier basis functions being shifted by the same $RV_0$.  In contrast, a signal varying its shape results in different basis functions being shifted by different amounts. \citet{jzhao2020} used the averaged effect of shifts in the lower and higher frequency ranges (denoted as $RV_\text{FT,L}$ and $RV_\text{FT,H}$ respectively) as a summary statistic to characterise the CCF behaviour. 

In this paper, we provide a more rigorous derivation of the \FIESTA methodology, and make updates to ease the interpretation of \FIESTA outputs.  For example, we abandon zero-padding (adding zeros at the end of the signal to increase sampling) as was used in the previous implementation of Fourier transform, since this acted as interpolating the power spectrum and the phase spectrum but did not add useful information. We also adopt the discrete Fourier transform (DFT), so that the output sampling size in the velocity frequency domain matches the input in the velocity domain, and thus for $N$ velocity grids in the CCF as input, there will be $N$ outputs in the corresponding ``velocity frequencies''. 

The paper is organised in the following manner. 
We provide an intuitive way to understand the maths behind the \FIESTA method and discuss its implementation in Section~\ref{sec:Method}. 
We discuss the practical consideration when applying \FIESTA to the parametrisation of CCFs and analysing timeseries in Section~\ref{sec:Measurement_uncertainties}. 
In Section~\ref{sec:FIESTA_SOAP}, we demonstrate the applications of \FIESTA on the SOAP~2.0 simulated solar spectra, with varying latitudes of solar spots and plages, as well as the simulated solar observation timeseries. 
Then it is followed by the applications of \FIESTA on the 3-years HARPS-N high-resolution solar observations in Section~\ref{sec:FIESTA_HARPS}.
A detailed comparison of \FIESTA models and similar models using more traditional activity indicators (FWHM and BIS) is carried out in Section~\ref{sec:BIS-FWHM}. 
Section~\ref{sec:discussions} is the summary and discussions where we also compare \FIESTA with previous works and discuss future research opportunities. 

Appendix~\ref{sec:coefficient_of_determination} defines the coefficient of determination $R^2$ and the adjusted $R^2$. 
Then in Appendix~\ref{sec:noise} we discuss the noise propagation in the Fourier transform and explore under what conditions the distributions of \FIESTA amplitudes and phases are nearly Gaussian.
Lastly, Appendix~\ref{appendix:zero-padding} demonstrates that zero-padding (implemented in the earlier version of \FIESTA) only changes the sampling of the Fourier transform output and so we do not adopt it anymore. 

%%%%%%%%%%%%%%%%%%%%%%%%%%%%%%%%%%%%%%%%%%%%%%%%%%%%%
\section{Method} 
\label{sec:Method}
%%%%%%%%%%%%%%%%%%%%%%%%%%%%%%%%%%%%%%%%%%%%%%%%%%%%%

%---------------------------------------------------
\subsection{\FIESTAtitle in a nutshell}
\label{subsec:FIESTA_nutshell}
%---------------------------------------------------
% Spectrographs used for Doppler planet surveys are typically designed to oversample the line spread function.
% %, and the smallest scale structure in the CCF is set by detector's pixel spacing.  
% Even though the wavelengths of detector pixels will typically not follow a linear grid, one can still evaluate the CCF along a regularly spaced set of velocities.
% Further, we evaluate the CCF of multiple observations at the same velocities relative to an inertial frame such as the solar system barycenter.
% %regardless of the motion of the Earth and observatory (i.e., changing barycentric correction).  

% CGTCGTCGT: I'm not sure what the point of the paragraph is. It is true
% that any two pixels don't over-sample the profile, the pixels not being 
% perfectly linear in velocity space, means that there IS sub-pixel sampling info
% available in the CCF when we look at the spectrum over many pixels and many
% orders AS A WHOLE. And then the whole thing lides against pixels over time
% due to earth. lotion.
%
% But this paragraph nevers draws any conclusion from these issues! What are you trying
% to say here?
%
\FIESTA is a Fourier decomposition of the CCF, whose flux $CCF(v_n)$ $(n=0, 1, \dots, N-1)$ are evaluated at a discrete grid of velocities, where the velocity grid $\{v_n\}$ is equally spaced and shared across observations.
\begin{equation}
	CCF(v_n) = \frac{1}{N}\sum_{k=0}^{N-1} \widehat{CCF}(\xi_k) e^{i\frac{2\pi}{N}nk} \qquad n=0, 1, \dots, N-1.
    \label{eq:CCF(v)}
\end{equation}
It is known as the inverse discrete Fourier transform (inverse DFT) that decomposes $CCF(v_n)$ into a linear combination of the orthogonal basis functions $e^{i\frac{2\pi}{N}nk}$, weighted by the complex coefficient $\widehat{CCF}(\xi_k)$.
Each coefficient $\widehat{CCF}(\xi_k)$ is derived from the DFT of $CCF(v_n)$
\begin{equation}
    \widehat{CCF}(\xi_k) = \sum_{n=0}^{N-1}CCF(v_n) e^{-i\frac{2\pi}{N}nk} \qquad k=0, 1, \dots, N-1.
\label{eq:DFT}
\end{equation}
The line shape information stored in $CCF(v_n)$ is lossless in the $\widehat{CCF}(\xi_k)$ if all $N$ terms are retained, as both can be converted via DFT and inverse DFT. 

We define, for each mode $k$ in the Fourier domain representation, the amplitude
\begin{equation}
    A_k = |\widehat{CCF}(\xi_k)|,
    \label{eq:A_k}
\end{equation}
the phase 
\begin{equation}
    \phi_k = \arctan{\frac{\widehat{CCF}(\xi_k)_\text{Im}}{\widehat{CCF}(\xi_k)_\text{Re}}}
    \label{eq:phi_k}
\end{equation}
and the velocity frequency
\begin{equation}
    \xi_k = k / L
    \label{eq:velocity_frequency}
\end{equation}
where $L$ the length of the CCF velocity grid and $1/L$ is the unit velocity frequency. Note that in this paper, the original domain is labeled in velocity $(v)$ and the Fourier transformed domain is labeled in ``velocity frequency'' ($\xi$). 
Because the inputs $CCF(v_n)$ are real functions and the DFT output is Hermitian-symmetric, only the first $N_0 = N/2+1$ or $(N+1)/2$ (depending on whether $N$ is even or odd) modes are needed for CCF parametrisation and the rest modes contain redundant information.

Replacing $\frac{nk}{N}$ by $\xi_k v_n$, we can rewrite the CCF in Eq.~\ref{eq:CCF(v)} in velocity $v_n$ as opposed to index $n$
\begin{equation}
	CCF(v_n) = \frac{1}{N} \sum_{k=0}^{N-1} \widehat{CCF}(\xi_k) e^{i 2\pi \xi_k v_n} \qquad n=0, 1, \dots, N-1.
    \label{eq:CCF(v)_freq}
\end{equation}
\FIESTA interprets the CCF shapes, as parametrised by the amplitudes $A_k$ and the phases $\phi_k$ at corresponding velocity frequencies $\xi_k$ with $k=0, 1, \dots, N_0-1$ in the Fourier domain.
In practice, we will keep only the leading terms which provide a dimensionally reduced representation of the CCF. The decision of how many terms to retain will depend on the spectral resolution, pixel sampling of the line spread function and SNR.  We discuss the choice for terms needed in Section~\ref{sec:Measurement_uncertainties}.
 
We use the python function \texttt{numpy.fft.rfft} to compute the DFT of the real input $CCF(v_n)$ with the Fast Fourier Transform algorithm \citep{CooleyTukey1965, press2007numerical}. 

%A continuous periodic function can be written as the sum of a constant mean and a linear combination of orthogonal sine and cosine terms, \jzhao{or a linear combination of the cosine terms parametrized by the amplitudes and phases if the sine terms are absorbed into the cosine terms.} Since the CCF is not periodic, we select a subset of the CCF near the systemic velocity and treat the result as one period of an infinite periodic function.

% In principle, the CCF can be computed as a continuous function of $v$, the velocity shift of the CCF mask.  
% However, the information content of the CCF as a function of $v$ is limited by both the instrument's spectral resolution and the pixelation of the sensor.  

%The \FIESTA outputs can be used to compute a reconstructed $\widehat{CCF}(v_n)$.    
%
% [{\bf TODO: Need to go through manuscript and differentiate between actual CCF and reconstructed CCF with widehat or some other notation.}]\jzhao{(keep original notation; the number of usable modes $k < N$ due to noise, but the decomposition remains valid.)}

%Note that $\hat{f}(\xi_k)$ is a complex number. We then compute the 
%In Appendix~\ref{sec:FS_vs_DFT}, we prove that the phases in Eq.~\ref{eq:CCF(v)_freq} and those calculated from the DFT in Eq.~\ref{eq:DFT} are identical and the amplitudes differ by a scaling factor $2/N$.

%---------------------------------------------------
\subsection{\FIESTAtitle applied to a pure line shift}
\label{subsec:line_shift}
%---------------------------------------------------
A pure radial velocity shift $RV$ due to orbiting exoplanets results in a bulk shift of the CCF as shown below.

\begin{align}
	CCF(v_n - RV) & = \frac{1}{N}\sum_{k=0}^{N-1} \widehat{CCF}(\xi_k) \cdot e^{i 2\pi \xi_k (v_n-RV)}
	\nonumber\\
	& = \frac{1}{N}\sum_{k=0}^{N-1} \widehat{CCF}(\xi_k) \cdot e^{i(2\pi \xi_k v_n - 2\pi \xi_k RV)}
\label{eq:CCF(v)_freq_shift}
\end{align}

Comparing Eq.~\ref{eq:CCF(v)_freq_shift} with Eq.~\ref{eq:CCF(v)_freq}, the $RV$ shift results in a phase shift $\Delta \phi_k = - 2\pi \xi_k RV$, which is consistent with the differential phase spectrum discussed in the continuous domain \citep{jzhao2020}. 
While the phase shift depends on the velocity frequency $\xi_k$, the ratio $\Delta \phi_k/\xi_k$ is invariant across velocity frequencies and proportional to the radial velocity shift.
\begin{equation}
    RV = -\frac{\Delta \phi_k}{2\pi \xi_k}.
\label{eq:RV_FIESTA_shift}    
\end{equation}
The amplitudes $A_k$ are also invariant for a pure line shift.

%---------------------------------------------------
\subsection{\FIESTAtitle applied to a perturbed line shape}
\label{subsec:line_shape}
%---------------------------------------------------
Substituting $\widehat{CCF}(\xi_k)$ by $A_k \cdot e^{i\phi_k}$ in Eq.~\ref{eq:CCF(v)_freq}, we have the following form
\begin{equation}
	CCF(v_n) = \frac{1}{N}\sum_{k=0}^{N-1} A_k \cdot e^{i\phi_k} \cdot e^{i 2\pi \xi_k v_n}.
    \label{eq:CCF(v)2}
\end{equation}
The shape of a deformed CCF can be characterised by changes in both the amplitudes $A_k$ and the phases $\phi_k$, and thus
\begin{align}
	CCF_\text{deformed}(v_n) & = \frac{1}{N}\sum_{k=0}^{N-1} (A_k+\Delta A_k) \cdot e^{i(\phi_k+\Delta \phi_k)} \cdot e^{i 2\pi \xi_k v_n}
	\nonumber\\
	& = \frac{1}{N}\sum_{k=0}^{N-1} (A_k+\Delta A_k) \cdot e^{i\phi_k} \cdot e^{i 2\pi \xi_k \big(v_n+\frac{\Delta \phi_k}{2\pi \xi_k}\big)}
    \label{eq:CCF(v)_freq_deform}
\end{align}

Comparing Eq.~\ref{eq:CCF(v)_freq_deform} with Eq.~\ref{eq:CCF(v)2}, the line deformations can be interpreted as an apparent (but spurious) radial velocity shift for each velocity frequency $k$.
%results in individual radial velocity shifts (with basis functions differentiated by the subscript $k$). 
%A Fourier basis with the velocity frequency $\xi_k$ is thus shifted by 
\begin{equation}
    RV_{\text{FT}, k} = -\frac{\Delta \phi_k}{2\pi \xi_k}.
\label{eq:RV_FIESTA_defo}        
\end{equation}
In addition to the phase-derived $RV_{\text{FT}, k}$, amplitude changes can also be used to trace a perturbed line shape.

% \ebf{I don't understand the choice of subscript FT.  There are Fourier transforms everywhere.  The distinguishing feature is that a perturbed line shape leads to an apparent/spurious radial velocity.  My first though is apparent for the subscript, but there might be a better choice.} \jzhao{be consistent with the FIESTA I}

%---------------------------------------------------
\subsection{Measuring \texorpdfstring{$\Delta \phi_k$}{Lg} and \texorpdfstring{$\Delta RV_k$}{Lg}}
%---------------------------------------------------
Measuring the phase shift $\Delta \phi_k$ requires specifying a reference CCF. In this manuscript, the weighted mean CCF serves as our phase reference. Throughout the paper (unless otherwise specified), we use the inverse variance weighting to calculate the weighted averages (e.g., mean CCF, daily binned RV, weighted RMS), where the variance is taken as the RV uncertainty squared.
Once we calculate $\Delta \phi_k$ between an observed and the reference CCF, Eq.~\ref{eq:RV_FIESTA_defo}
provides a group of relative radial velocity shifts $RV_{\text{FT}, k}$ between the two CCFs for each velocity frequency $\xi_k$. 
Similar to $RV_{\text{FT,L}}$ and $RV_{\text{FT,H}}$ proposed in \cite{jzhao2020} that represent the RV shifts in the lower and higher frequencies, $RV_{\text{FT}, k}\,(k=0, 1, \dots, N_0-1)$ will show the same response to a bulk line shift but different responses to line shape changes. 
Therefore,
\begin{equation}
    \Delta RV_k \equiv RV_{\text{FT}, k} - RV_{\text{apparent}} \qquad k=0, 1, \dots, N_0-1
\label{eq:delta_RV_k}
\end{equation}
is used to parametrise the line profile deformation, where $RV_{\text{apparent}}$ is a single apparent RV shift derived from the full CCF.
$RV_{\text{apparent}}$ may be computed using an independent RV measurement algorithm or as a weighted mean of $RV_{\text{FT},k}$'s that account for varying measurement precision (e.g. Section~\ref{sec:weighted_mean}).
In this paper, we use $RV_{\text{Gaussian}}$, which is the centre of a Gaussian fit to the CCF extending to $5\sigma$ in either direction from centre.

\subsection{Summary}
The effects of a pure shift (Eq.~\ref{eq:RV_FIESTA_shift}) and the effects of line shape changes (Eq.~\ref{eq:RV_FIESTA_defo}) are fundamentally different,  even though the expressions are similar. The $RV$ in Eq.~\ref{eq:RV_FIESTA_shift} is the same for all velocity frequencies $\xi_k$, whereas the $RV_{\text{FT}, k}$ in Eq.~\ref{eq:RV_FIESTA_defo} changes with the Fourier  mode $k$. 
We treat Eq.~\ref{eq:RV_FIESTA_shift} as a special case of Eq.~\ref{eq:RV_FIESTA_defo} where $RV_{\text{FT}, k}$ is a constant for all $k\,(k=0, 1, \dots, N_0-1)$. 
As a result, we can now use a group of $RV_{\text{FT}, k}\,(k=0, 1, \dots, N_0-1)$ to parametrise the CCF behaviour (be it a bulk shift, a deformation, or both).
While a single measurement of the apparent radial velocity shift per observation can not distinguish between the effects of orbiting exoplanets and stellar variability, the set of $RV_{\text{FT}, k}$'s can be used to recognise the presence of line shape changes. We explore that possibility using simulated and actual observations in Section~\ref{sec:FIESTA_SOAP} and Section~\ref{sec:FIESTA_HARPS}.  

Changes in the Fourier amplitudes ($A_k$) also indicate line variability. However, CCF shape changes in the vertical direction that may not induce apparent RV variation can also impact $A_k$. In this paper, as we intend to model the line variability induced RV, we focus on the analysis of $RV_{\text{FT}, k}$ and $\Delta RV_k$, which are directly related to the spurious RV (see Section~\ref{sec:weighted_mean} for further discussions).

%%%%%%%%%%%%%%%%%%%%%%%%%%%%%%%%%%%%%%%%%%%%%%%%%%%%%
\section{Useful Fourier  modes in \FIESTAtitle}
\label{sec:Measurement_uncertainties}
%%%%%%%%%%%%%%%%%%%%%%%%%%%%%%%%%%%%%%%%%%%%%%%%%%%%%

Noise in the CCF propagates into the uncertainties in the \FIESTA amplitudes and phase (and thus radial velocity shifts calculated from Eq.~\ref{eq:RV_FIESTA_defo}). We need to understand the effects of noise in order to interpret the results of the \FIESTA analysis. In principle, using all $N$ Fourier modes (or $N_0$ Fourier modes for real inputs as in our case) can reconstruct the CCF without losing information. In practice, measurement noise and/or the spectrograph resolution may limit the utility of higher frequencies.  In this section, we consider each of the factors that may limit the number of useful terms in the \FIESTA reconstruction.
We summarise five aspects to consider when it comes to making use of the amplitudes $A_k$ and phases $\phi_k$ information in the presence of associated uncertainties in the Fourier domain as below. 
\begin{enumerate}
    %\item Do $A_k$ and $\phi_k$ represent intrinsic changes in the CCF or photon noise? It can be an issue for higher $k$'s.
    %\item Are the uncertainties correctly represented in the Fourier domain, i.e whether they follow a normal distribution so that the $1\sigma$ uncertainty can be used?
    \item How many Fourier modes are necessary to accurately and adequately reconstruct the typical CCF shape?
    \item What is the distribution of errors in $A_k$ and $\phi_k$ due to photon noise?
    \item From which $k$ does the SNR of $A_k$ and $\phi_k$ become so low that the estimates of individual $A_k$ and $\phi_k$ are no longer useful?
    % \item Do estimates of the $A_k$'s and $\phi_k$'s have enough precision to characterise deviations of the CCF at each epoch from the template CCF?
    % \item Looking at the timeseries, at which $k$ do the variations of $A_k$ and $\phi_k$ over time become insignificant compared to the errors of $k$ the $A_k$?
    \item From which $k$ is the variability in the $A_k$ and $\phi_k$ timeseries dominated by the measurement uncertainty of $A_k$'s and $\phi_k$'s?
    \item Which terms of the \FIESTA decomposition correspond to velocity scales greater than the width of the spectrograph line spread function (LSF)? 
\end{enumerate}
They are discussed in the following subsections respectively.

%---------------------------------------------------
\subsection{Reconstructing the CCF}
\label{sec:CCF_decomposition}
%---------------------------------------------------

First, we consider how many Fourier modes are needed for \FIESTA to accurately parametrise a typical CCF without overfitting.
The number of modes needed depends on the shape of the CCF and the modelling error, which is defined as the residual root mean square (RMS) between the observed CCF and the \FIESTA reconstruction of the CCF using only the first $k$ terms. Too few terms would result in an inaccurate representation of the CCF due to a large residual, but too many terms would result in a modelling error below the photon noise, indicating overfitting.

\begin{figure*}[t!]
\centering
    \includegraphics[width=1\columnwidth]{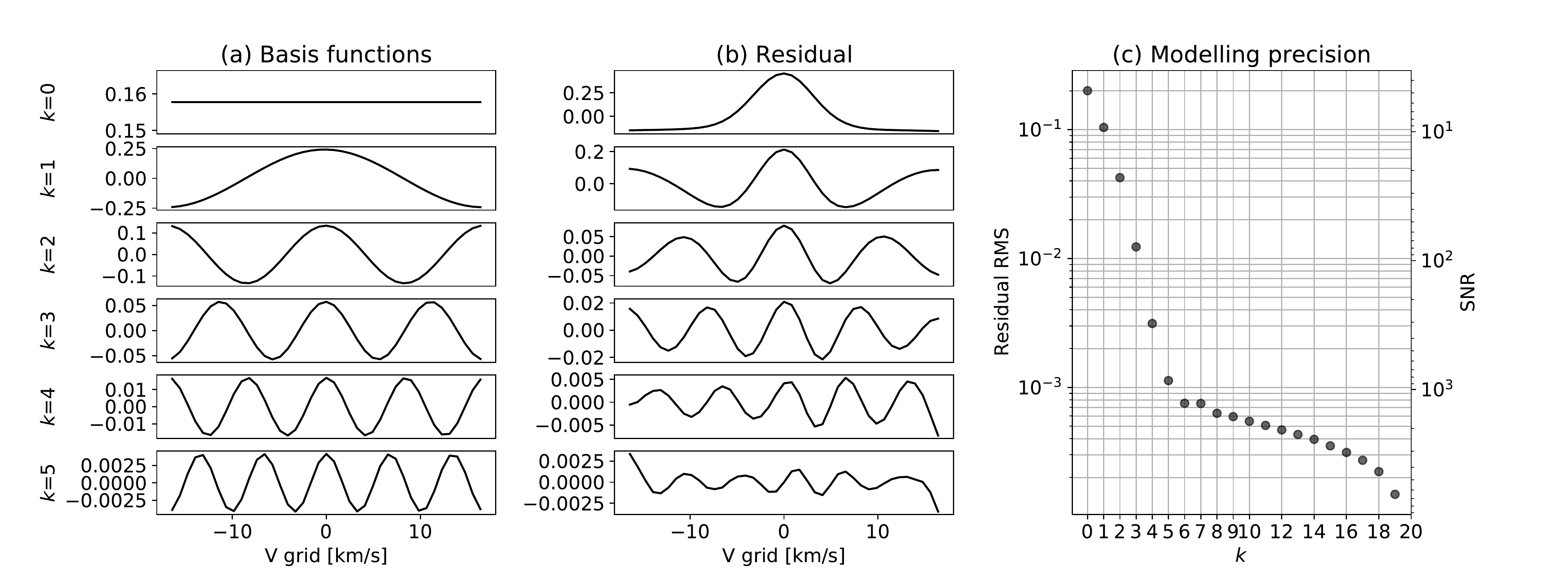}
    \caption{
    (a) The \FIESTA basis functions used to fit a HARPS-N solar spectrum CCF up to $k=5$ Fourier modes. 
    (b) The residual of the CCF after fitting the first $k$ Fourier modes. 
    (c) The modelling precision (i.e. residual RMS) as a function of the number of modes up to $k$ used to parametrise the CCF. The vertical axis on the right stands for the SNR to which the CCF can be reconstructed within the photon noise uncertainties. Using 20 Fourier modes will fully recover the CCF (i.e., zero modelling noise; not shown). The modelling precision as a function of $k$ is for illustration only and it may vary if a different CCF profile is used, e.g., from a different star or a different instrument LSF.
    }
    \label{fig:Flux_residual}
\end{figure*}

For example, Fig.~\ref{fig:Flux_residual} shows the first 5 \FIESTA basis functions (or 6 including the zeroth mode, which is the mean function of the CCF) used to fit an arbitrary but typical HARPS-N solar spectrum CCF. For details of the solar spectrum, refer to \cite{Dumusque2021HARPS-N} and Section~\ref{sec:FIESTA_HARPS}. 
Using up to $k=5$ \FIESTA basis functions reaches the residual or modelling error of $6.7\times10^{-4}$, corresponding to a SNR\footnote{It is approximately calculated as the median flux of the CCF between -15 and 15 km/s (signal) divided by the modelling error (noise).}~=~$1.4\times10^3$ for the CCF it is able to represent.
The HARPS-N solar CCFs have a high median SNR $\sim$ 10,000 for individual CCFs and SNR $\sim$ 46,000 for the daily binned CCFs. In order for the modelling error to be less than the photon noise level, one would need to use all the available Fourier modes, which is $k_\textrm{max,photon} = 20$. Note that $k_\textrm{max,photon}$ is recommended for extracting all usable information of the CCF up to the photon noise limit, but any $k\le k_\textrm{max,photon}$ provides valuable information.

% The right hand axes of Fig.~\ref{fig:Flux_residual} gives the minimum SNR required in order for the photon noise to be less than the residual between the CCF and the \FIESTA reconstruction of the CCF using only $k$ terms.  
% For example, with $k=6$, the minimum CCF SNR would be 1446, significantly less than the typical SNR of HARPS-N solar CCFs.  
 
%Therefore, for the HARPS-N solar observations, it will be important to use at least 6 $A_k$'s and $\phi_k$'s to represent the shape of the the CCF up to the photon noise modelling error. 

%-------------
%-------------

%---------------------------------------------------
\subsection{Noise distribution of amplitudes and phases}
\label{sec:noise_distribution}
%---------------------------------------------------

According to Eq.~\ref{eq:DFT}, $\widehat{CCF}(\xi_k)_{\mathrm{Re}}$ and $\widehat{CCF}(\xi_k)_{\mathrm{Im}}$ can be obtained by a linear projection of $CCF(v_n)$ onto a set of Fourier basis vectors.  
Therefore, Gaussian noise in $CCF(v_n)$ translates into Gaussian errors for $\widehat{CCF}(\xi_k)_{\mathrm{Re}}$ and 
$\widehat{CCF}(\xi_k)_{\mathrm{Im}}$. 
In fact, as derived in Appendix~\ref{sec:noise}, their $1\sigma$ uncertainties are
\begin{equation}
    \sigma_{\widehat{CCF}(\xi_k)_\text{RE}} = \sigma_{\widehat{CCF}(\xi_k)_\text{IM}} = 
    \sqrt{\frac{1}{2} \sum_{n=0}^{N-1} \sigma^2_{CCF(v_n)}} \coloneqq
    \widehat{\sigma}
    \label{eq:sigma_re_im}
\end{equation}
where $\sigma_{CCF(v_n)}$ is the uncertainty of $CCF(v_n)$.

In contrast, the amplitude,  phase and the resulting $RV_{\text{FT}, k}$ are obtained by non-linear transformations (i.e., square and square root operations on two random variable in Eq.~\ref{eq:A_k} and the ratio and arctan operations in Eq.~\ref{eq:phi_k}). Therefore, the uncertainties in $A_k$'s, $\phi_k$'s and $RV_{\text{FT}, k}$'s are not strictly Gaussian. 
The amplitudes tend to be skewed with a lower boundary at zero and the phases tend to present a narrower core and broader tails (Fig.~\ref{fig:normality_test}). These effects can be substantial when the input CCFs data are noisy. 
We define the noise level (NL) as
\begin{equation}
    \text{NL}
    =
    \frac{\widehat{\sigma}}{A_k}.
    \label{eq:CCF_signal_noise}
\end{equation}
Based on numerical simulations, we have verified that noise in measurement of amplitudes and phases follow nearly the normal distribution when NL~$\leq$~0.2 (Appendix~\ref{sec:noise}).
Under this low NL regime, we can approximately derive the analytical expression for the uncertainties of amplitudes and phases
\begin{equation}
    \sigma_{A_k} = \widehat{\sigma} 
    \quad
    \text{and}
    \quad
    \sigma_{\phi_k} = \frac{\widehat{\sigma}}{A_k}.
    \label{eq:sigma_A_phi}
\end{equation}
In general, the higher the mode $k$ becomes, the smaller the amplitudes and thus the larger the NL. 
We define a cut-off frequency index based on normality $k_\text{max,normal}$ and $\xi_{k_\text{max,normal}}$ as the last frequency that satisfies our criterion.
%We compute the uncertainties $A_k$'s and $\phi_k$'s based on propagating measurement uncertainties from the CCF (Eq.~\ref{eq:sigma_A_phi}). 
By limiting our analysis to $k \le k_{\max,\text{normal}}$, we ensure that noise in $A_k$ and $\phi_k$ closely follows a normal distribution.
%While a ratio of NL~=~0.4 could be adopted to include a larger range of Fourier  modes without introducing much bias on the uncertainty estimation (Appendix~\ref{sec:noise}), this is unnecessary for the observations analysed in this paper.

%---------------------------------------------------
\subsection{Individual \texorpdfstring{$A_k$ and $\phi_k$}{Lg} SNR}
\label{sec:cut-off frequency1}
%---------------------------------------------------
For an individual observation, we compute the SNR for $A_k$'s and $\phi_k$'s as $A_k/\sigma_{A_k}$ and $1/\sigma_{\phi_k}$ \footnote{As the phase $\phi_k$ ranges between $-\pi$ and $\pi$, a larger $|\phi_k|$ does not necessarily mean the signal is stronger. We therefore use the unity instead of $|\phi_k|$ in the numerator to calculate the phase SNR.}
using the uncertainties from that observation. 
The SNR for $A_k$'s and $\phi_k$'s are thus identical according to Eq.~\ref{eq:sigma_A_phi}.
As $A_k$ decreases rapidly while $\sigma_{A_k}$ is nearly a constant (see Eq.~\ref{eq:sigma_re_im} and \ref{eq:sigma_A_phi}) with increasing $k$, the SNR for $A_k$'s and $\phi_k$'s decreases for higher $k$.
Therefore, we limit our analysis to $k\le k_{\max,\textrm{individual}}$, where $k_{\max,\textrm{individual}}$ is the largest $k$ for which the median SNR of $A_k$'s or $\phi_k$'s is at least 2.

When choosing which $k$ to analyse in a timeseries, we use the median $\sigma_{A_k}$ and median $\sigma_{\phi_k}$ instead of the uncertainty for each observation, so as to ensure consistency across observations and to avoid discarding information in $k$-th timeseries that would result in only a small fraction of low SNR observations not passing our minimum SNR threshold.  
%
%we would require that the SNR for $A_k$'s and $\phi$'s exceed our minimum SNR threshold.   
%. Ideally, we want to be confident about each of the measured $A_k$ and $\phi_k$ by setting a minimum SNR (i.e. $A_k/\sigma_{A_k}$ and $\phi_k/\sigma_{\phi_k}$), thus requiring all the $A_k$ and $\phi_k$ in the timeseries pass the minimum SNR. 
%However, it is not advisable to throw away the whole timeseries because a small fraction of the observations have relatively low SNR. 
%Therefore, we adopt to use the median SNR for each timeseries. 

%As SNR for $A_k$'s and $\phi_l$'s decreases for higher $k$, we decide once the median SNR~$<2$, the timeseries will be abandoned. %We denote this cut-off frequency as $\xi_\text{individual}$.

%---------------------------------------------------
\subsection{Timeseries SNR}
% \subsection{SNR of \texorpdfstring{$A_k$ and $\phi_k$}{Lg} variations with time}
\label{sec:cut-off frequency2}
%---------------------------------------------------

When applying \FIESTA to a timeseries of CCFs, we aim to characterise the deviations of ${A_k}$ and $\phi_k$ from their time-averages.  
We define a $\mathrm{SNR}_{\mathrm{timeseries},k}$ as the ratio of the RMS deviation of each $A_k$ and $\phi_k$ from their mean value over a timeseries to the median $\sigma_{A_k}$ and $\sigma_{\phi_k}$. 
%
%If a star shows no variability and the instrument is ultra stable, then the only variability in $A_k$'s and $\phi_k$'s among observations is only due to photon noise. In this hypothetical scenario, there won't be any significance of the variation of $A_k$ and $\phi_k$ timeseries against the overall $\sigma_{A_k}$ and $\sigma_{\phi_k}$. In other words, the $A_k$ and $\phi_k$ timeseries could be modeled as constants plus measurement noise. 
%A similar situation happens when analysing varying CCFs in higher Fourier  modes. 
%Most of the information in $A_k$'s and $\phi_k$'s is concentrated in low $k$'s, since %
%As $k$ increases, it can reach a point where 
%the overall $\sigma_{A_k}$ and $\sigma_{\phi_k}$ will exceed the variations of the $A_k$ and $\phi_k$ timeseries for large $k$.
%, making the timeseries insignificant from being varying.

% When applying \FIESTA to a timeseries of CCFs, we aim to characterise the deviations of the CCF from the time-averaged CCFs. 
%
It is useful to limit our analysis to the $A_k$'s and $\phi_k$'s that have measurement uncertainties less than the extent of their variations with time.  
In this paper, we require a minimum SNR of 2 to ensure high-quality timeseries are obtained, i.e., the standard deviation of an $A_k$ or $\phi_k$ timeseries is at least twice as large as the median $\sigma_{A_k}$ or median $\sigma_{\phi_k}$ in the timeseries. 
We define $k_{\max,\text{timeseries}}$ as the maximum $k$ for which this criterion is satisfied.  
 
%Generally speaking, the uncertainties of $A_k$ and $\phi_k$ become higher in higher Fourier  modes. When it come to the timeseries of the Fourier amplitudes $A_k$ and phases $\phi_k$, it can reach a point where the overall uncertainties of $A_k$ and $\phi_k$ exceed the variations of the $A_k$ and $\phi_k$ timeseries, making the timeseries insignificant from being a constant signal. An extreme case would be observing a completely quiet star with ultra stabilised instrument - the variation in $A_k$ and $\phi_k$ timeseries would be simply due to the photon noise. Therefore, a cut-off frequency $\xi_\text{timeseries}$ for all the CCFs is needed to make sure the timeseries has significant enough SNR to reveal the variations of the observations (be it from the star or the instrument), where the signal is the variation of the timeseries and the noise is the overall uncertainty based on each measurement. In this paper, we choose SNR = 2 as the minimum SNR for all timeseries to ensure high-quality timeseries are obtained, i.e., the standard deviation of a timeseries is at least twice as large as the median uncertainty of individual measurement in the $A_k$ and $\phi_k$ timeseries. 

%---------------------------------------------------
\subsection{Instrumental resolution}
%---------------------------------------------------
Finally, we consider $k_{\max,\mathrm{inst}}$, the maximum $k$ for which we expect changes in the CCF could be dominated by the target star, rather than the instrument or detector.
%The last note to consider would be the Fourier  modes that reflect changes in the star rather than the instrument. 
Due to the instrumental broadening of the spectral lines, higher Fourier  modes that captures changes below the LSF width will not be useful in parametrising stellar variability. 
For example, the instrumental resolution of HARPS in velocity space is 2.5~km/s.
%\footnote{Resolution in velocity = speed of light (3$\times10^8$~m/s) / spectral resolution (120,000).} 
For a CCF with a velocity ranging from $-15$ to 15~km/s, Fourier  modes higher than $k=L/(1/\xi_k) = 30/2.5 =12$ (Eq.~\ref{eq:velocity_frequency}) capture variations on a scale finer than  the instrumental resolution.
%element at 2.5~km/s, which might just be instrumental instability. 
In this example, we limit the $A_k$'s and $\phi_k$'s analysed to have $k \le k_{\max,\mathrm{inst}} = 12$. As shown in Section~\ref{sec:FIESTA_HARPS}, the solar rationally induced variability are barely seen in modes with $k>12$.

\subsection{Summary notes}
\label{sec:Summary_notes}
For illustrative purposes, we use the three years of HARPS-N solar CCFs (details in Section~\ref{sec:FIESTA_HARPS}) to give readers an idea of what $k$ is recommended based on the criteria discussed above.
\begin{equation*}
    k = \min(k_\textrm{max,photon}, k_\text{max,normal}, k_{\max,\textrm{individual}}, k_{\max,\text{timeseries}}, k_{\max,\mathrm{inst}}) = \min(20, 20, 20, 11, 12) = 11
\end{equation*}
will ensure that the $A_k$'s and $\phi_k$'s being analysed can be associated with an unbiased $1\sigma$ uncertainty and have sufficient SNR to characterise the CCF and the timeseries variations as well as focusing on revealing information that could be due to stellar variations (as opposed to instrumental or detector variations).
%For HARPS-N solar spectra, the maximum $k$ is set by the SNR of the timeseries $k_{\max,\text{timeseries}}$ (\S reference), as the sun is a relatively quiet star.}
%
While \FIESTA is capable of including higher $k$'s, its default choice of thresholds (and those used in this study) are designed to be conservative in ensuring that the measurement uncertainties are very nearly Gaussian.
%Overall, the SNR used to determine the cut-off frequency should depend on whether we emphasize high-quality timeseries by sacrificing some higher frequencies, or prefer the completeness of Fourier  modes while keeping the lower SNR timeseries. \FIESTA provides these cut-off frequencies based on the user input (e.g. SNR) and allows users to decide how many Fourier modes to be used. \jzhao{a comparison among $\xi_\text{normal}$, $\xi_\text{individual}$ and $\xi_\text{timeseries}$}
% For the cases that we have studied so far, SNR = 2 for the timeseries is usually a stronger condition and restricts the cut-off frequency further than what was discussed in Section~\ref{sec:Measurement_uncertainties}

%%%%%%%%%%%%%%%%%%%%%%%%%%%%%%%%%%%%%%%%%%%%%%%%%%%%%
\section{Validation of \FIESTAtitle on simulated solar observations}
\label{sec:FIESTA_SOAP}
%%%%%%%%%%%%%%%%%%%%%%%%%%%%%%%%%%%%%%%%%%%%%%%%%%%%%
We use the SOAP~2.0 (Spot Oscillation And Planet 2.0, \citealp{Dumusque2014SOAP}) simulator  to generate CCFs of solar spectra affected by spots or plages that are based on the HARPS observations of the Sun. 
Unless specified in this paper, we set the spot at $30^\circ$ latitude with the size of $0.1 R_\sun$; the stellar configurations are based on the Sun and chosen as the default values of SOAP~2.0, e.g. $T_\text{star}=5778~K$, $T_\text{spot-photosphere}=663~K$.

% As was discussed in \cite{jzhao2020}, the higher the frequency $\xi_k$, the phase measurement is more prone to affected by photon noise because the amplitudes in higher frequencies are smaller. To identify the cut-off frequency $\xi_k'$ above which the basis functions become noise-dominated, we seek $\xi_k'$ such that abandoning information with $\xi_k > \xi_k'$ (i.e. setting $\hat{f}(\xi_k)$ with $\xi_k > \xi_k'$ to zeros) would result in the difference between its inverse Fourier transform and the CCF smaller than the photon noise level. Fig.~\ref{fig:Flux_residual} shows how such differences decrease as frequencies used to recover a \jzhao{simulated Gaussian
% $CCF(v) = \frac{1}{{2 \sqrt {2\pi } }}e^{{{ - {x }^2 } \mathord{\left/ {\vphantom {{ - {v} ^2 } {2 \times 2 ^2 }}} \right. \kern-\nulldelimiterspace} {8 }}}$}
% via the inverse Fourier transform increases. 

%---------------------------------------------------
\subsection{Line shift vs line deformation}
\label{sec:challenging_test}
%---------------------------------------------------
First we generate the CCFs of solar spectra affected by a single spot in one rotation. In the presence of stellar variability, each $RV_{\text{FT}, k}$ differs for each CCF, because each mode $k$ traces the line deformation at a different length-scale, with $k=1$ tracing variability at the full range of the CCF, $k=2$ tracing variability at 1/2 the range, etc. Therefore, $RV_{\text{FT}, k}$ provides a quantitative measurement of the radial velocity shift and the line deformation.

\begin{figure}[t]
    \includegraphics[width=1\columnwidth]{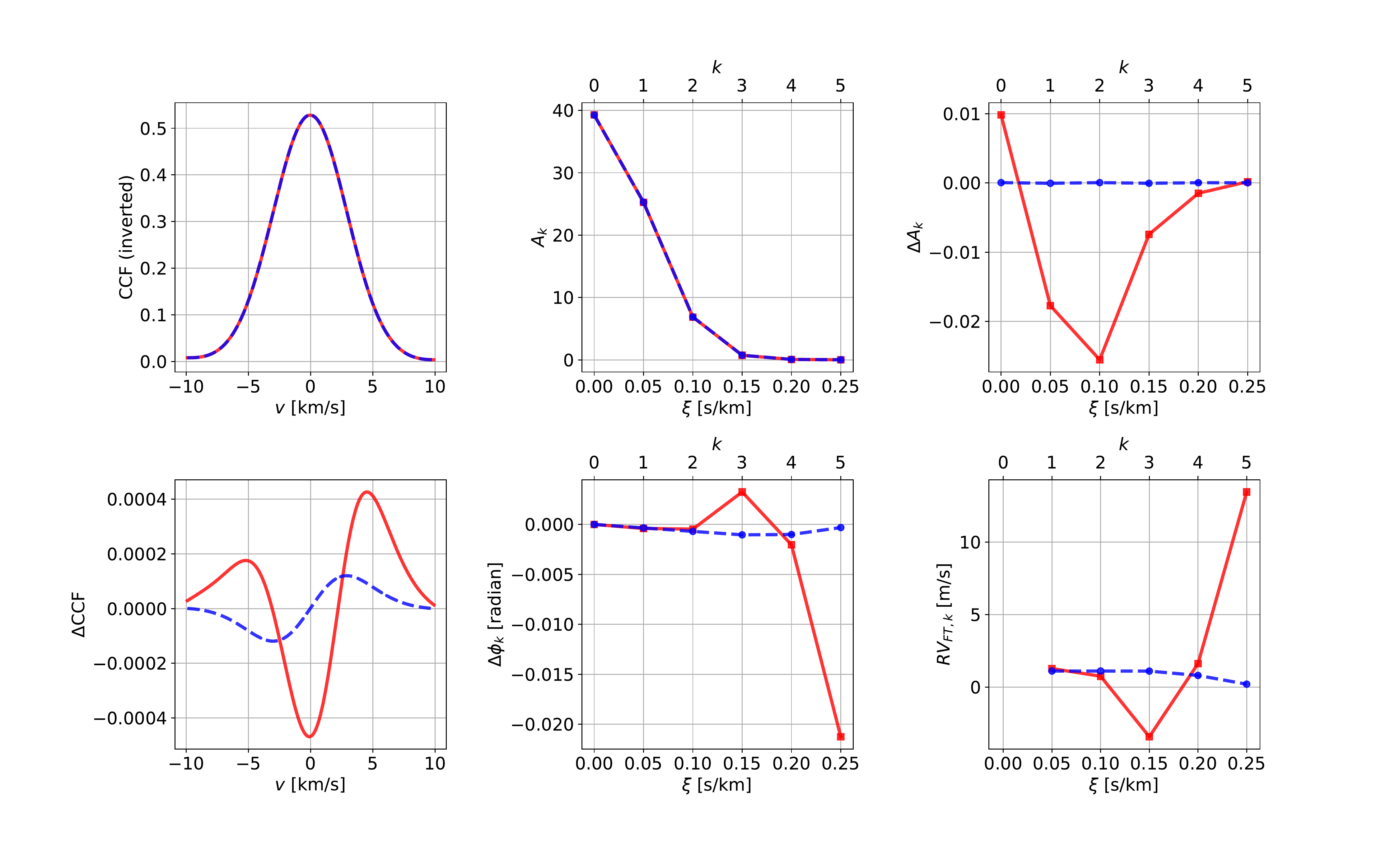}
    \caption{Amplitudes and phases at a glance for two CCFs, one caused by a pure shift (blue dashed line) and the other caused by a spot at $30^\circ$ latitude with the size of $0.1 R_\sun$ at rotation phase 0.51 (red solid line). $\Delta\textrm{CCF}$, $\Delta A_k$, $\Delta \phi_k$ and $\Delta RV_k$ are relative measurements to the template CCF which is neither shifted or deformed. Even though both RVs are comparable (1.1~m/s), \FIESTA tells them apart (one is intrinsic and the other is spurious) and provide a quantitative measurement of the phase and amplitude changes.
    Note that it is impossible to replicate a line profile with a shift below the sampling spacing, and so even the pure shift of a CCF simulated by interpolation will end up in a slightly different shape and result in unequal RV shifts in higher frequencies when the amplitudes are smaller.}
\label{fig:challenging test}
\end{figure}

To illustrate the changes in the Fourier amplitudes and phases of the CCFs for a line deformation as opposed to a line shift, we take one snapshot at solar rotation phase 0.51, when the spot is slightly off solar disk centre and the resulting CCF is deformed with an apparent radial velocity measured at 1.1\,m/s. For comparison, we shift the undeformed CCF by the same 1.1\,m/s and feed them into \fiesta. When the CCF is decomposed into the Fourier basis functions, the amplitudes and RV shifts remain the same for all modes for a pure line shift, while amplitudes and RV shifts vary for a line deformation (Fig.~\ref{fig:challenging test}). 

%-------------
%-------------

%---------------------------------------------------
\subsection{Weighted mean of \texorpdfstring{$RV_{\text{FT}, k}$}{Lg}}
\label{sec:weighted_mean}
\begin{figure*}[t!]
\centering
    \includegraphics[width=0.49\columnwidth]{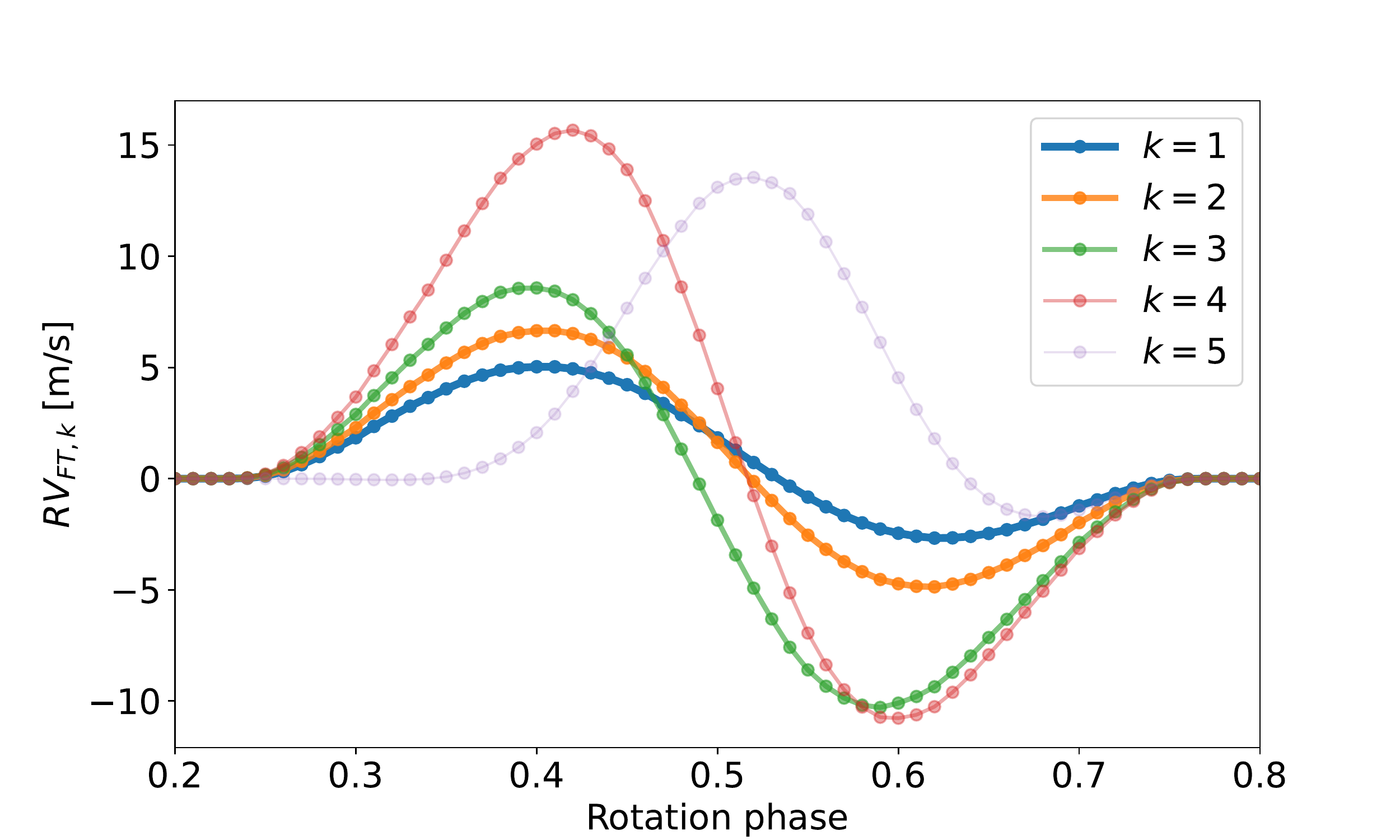}
    \includegraphics[width=0.49\columnwidth]{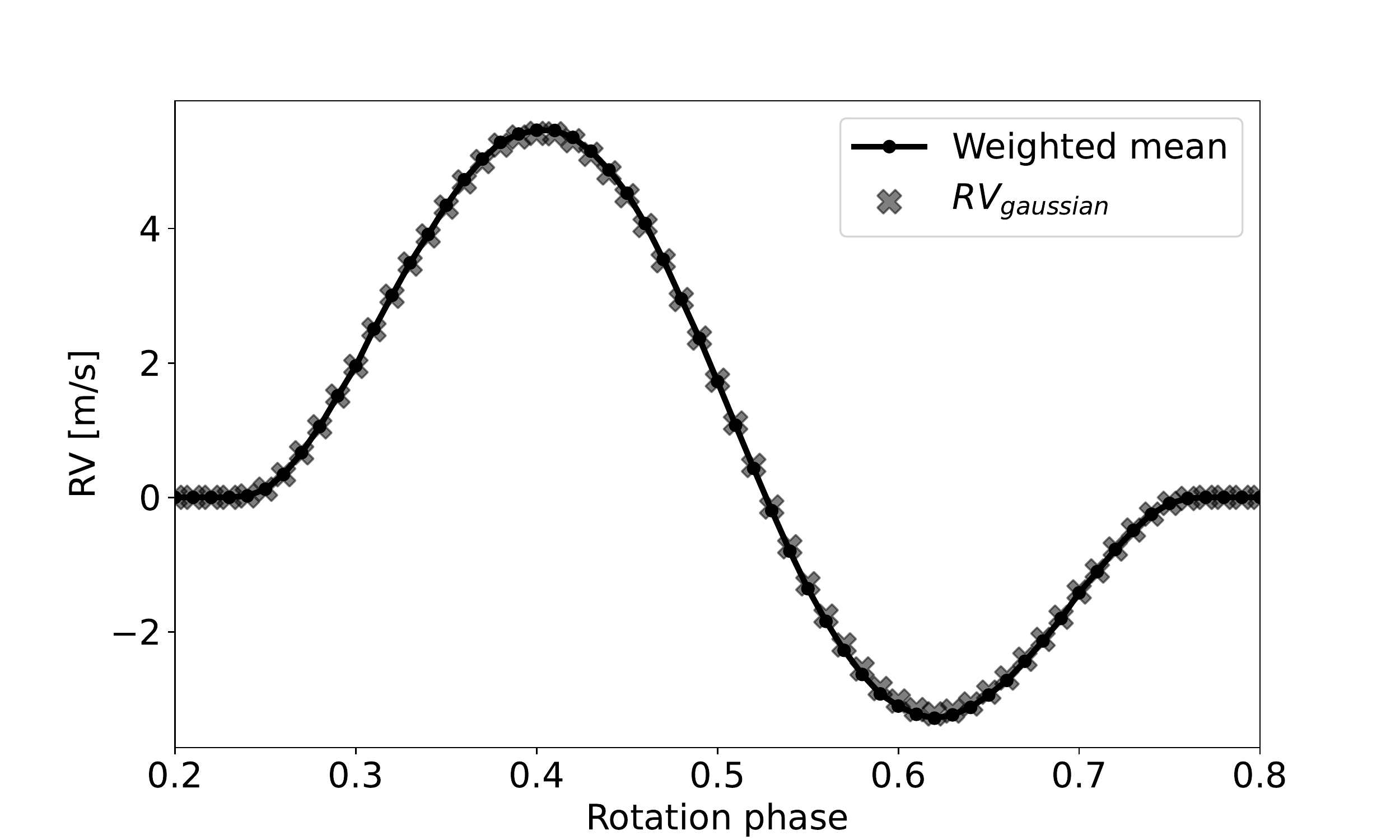}
    \caption{Left panel: timeseries of the first 5 $RV_{\text{FT}, k}$ over the course of solar rotation for SOAP~2.0 simulations of a single solar spot of $0.1 R_\sun$ at $30^\circ$ latitude. Only rotation phases between 0.2 and 0.8 are shown. Right panel: the weighted mean of $RV_{\text{FT}, k}$ overplotted with $RV_{\text{Gaussian}}$. It shows while individual $RV_{\text{FT}, k}$ traces the line deformation differently, the overall effect of $RV_{\text{FT}, k}$ is consistent with $RV_{\text{Gaussian}}$, indicating the set of $RV_{\text{FT}, k}$ is a multi-dimension measurement of the RV shift that has an overall effect equivalent to $RV_{\text{Gaussian}}$.
    % \ebf{Need to say what data is being analysed.  What do you want reader to take away from the figure?}
    }
    \label{fig:RV_insight}
\end{figure*}

%---------------------------------------------------
We calculate the weighted mean of $RV_{\text{FT}, k}$ for the SOAP simulated CCFs over the course of one rotation, with the weights being the corresponding amplitudes $A_k$. We find that the weighted means are consistent with the radial velocity shifts measured by fitting a Gaussian function to the CCF (Fig.~\ref{fig:RV_insight}). Therefore, the apparent radial velocity $RV_{\text{Gaussian}}$ can be treated as the overall effect of the radial velocity shifts of individual Fourier basis functions (i.e. $RV_{\text{FT}, k}$). $RV_{\text{FT}, k}$, on the other hand, can be treated as a multi-dimensional measurement of the RV shift at different CCF length scales.
%-------------
%-------------

%---------------------------------------------------
\subsection{Spots vs plages and their \FIESTAtitle correlations}
\label{sec:spot_plage_correlations}
%---------------------------------------------------
We further explore how \FIESTA behaves under the influence of a solar spot and plage in various latitudes (from 10 to 80 degrees) in one stellar rotation. 
Earlier studies using the previous version of \FIESTA can be referred to \citet{jzhao_thesis}.

\begin{figure*}[t]
\centering
    \includegraphics[width=0.49\columnwidth]{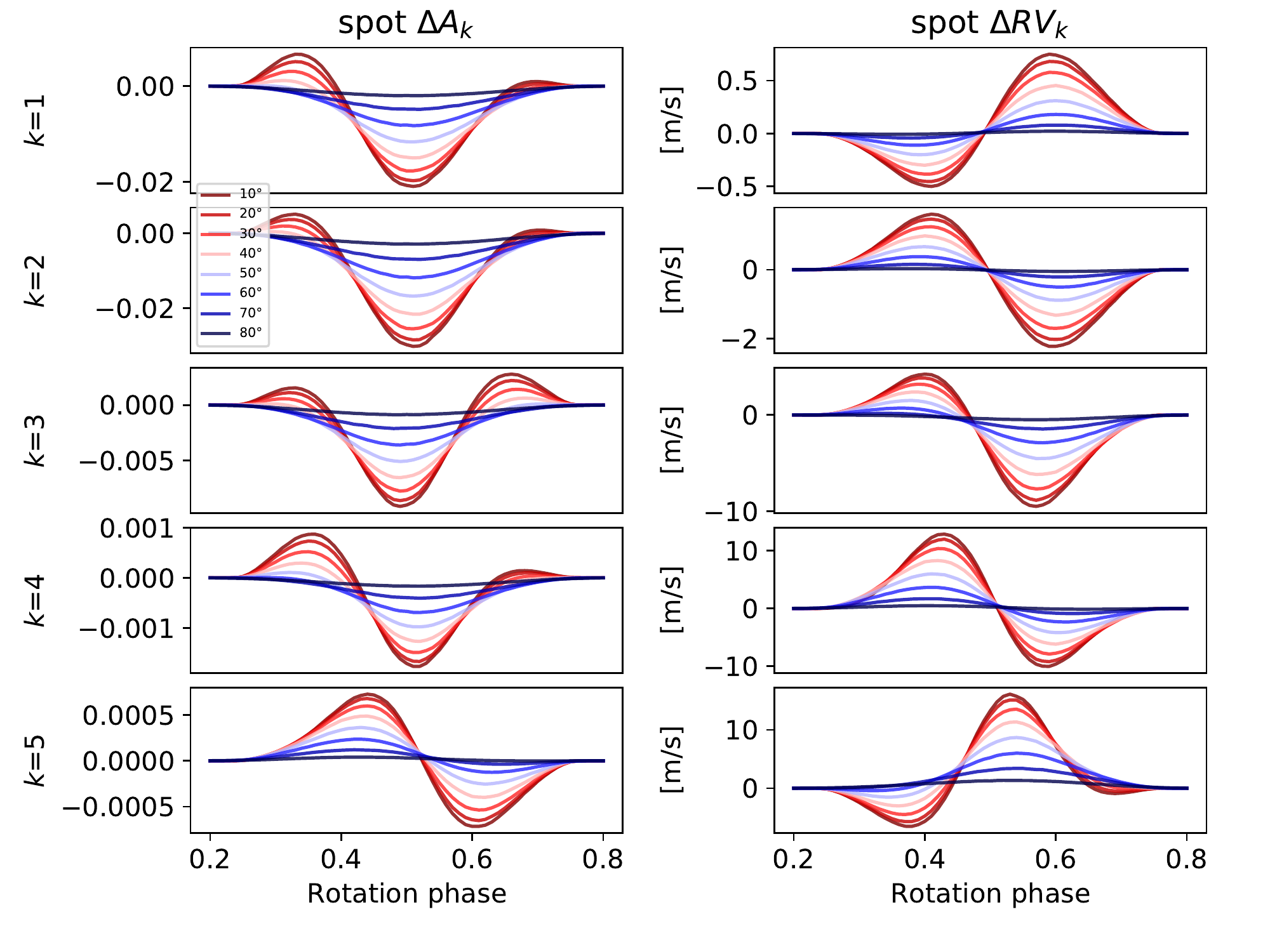}
    \includegraphics[width=0.49\columnwidth]{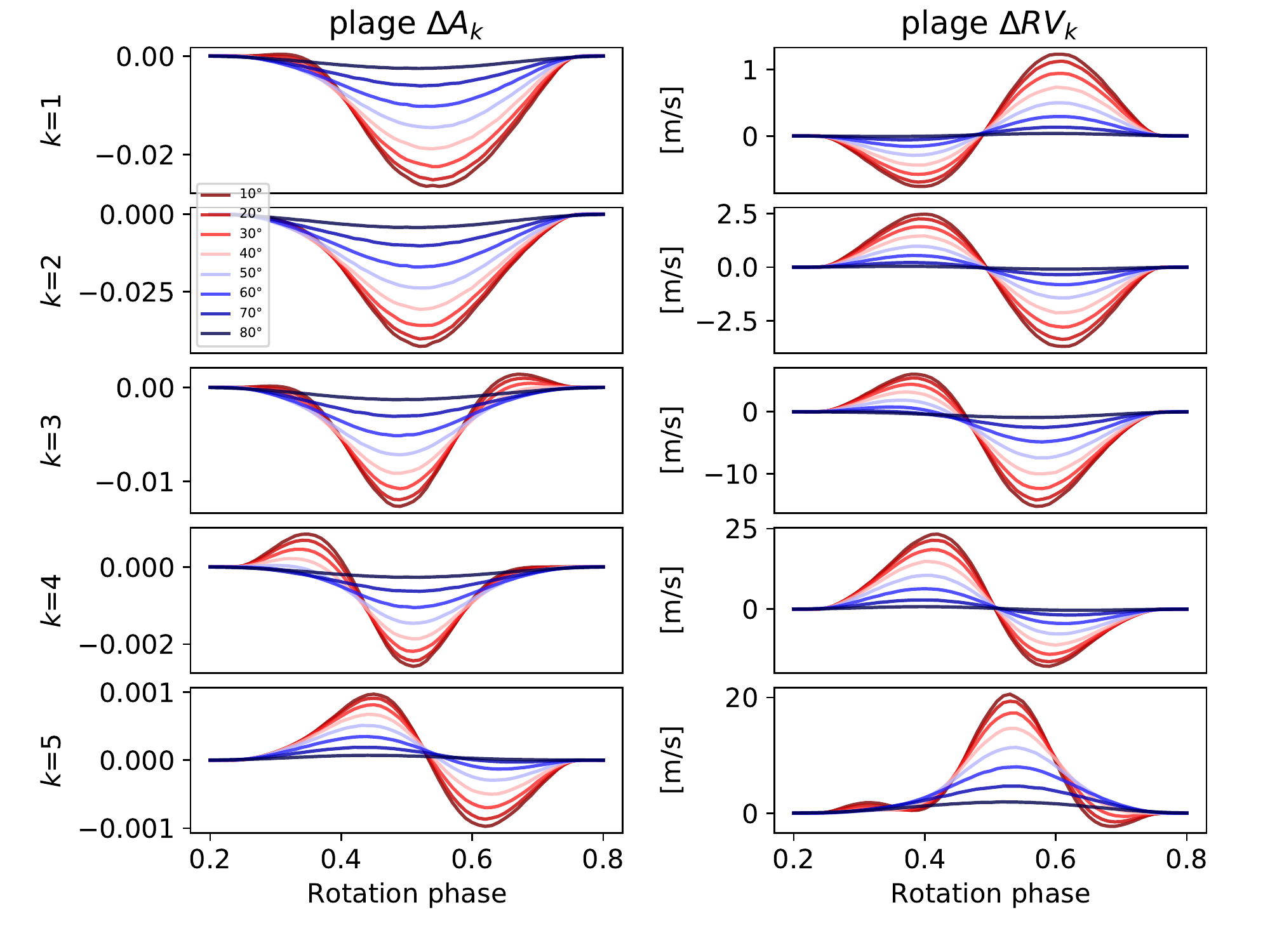}
    \caption{$\Delta A_k$ and $\Delta RV_k$ in one solar rotation for spots (left two panels) and plages (right two panels), simulated in latitudes from 10 to 80 degrees.}
    \label{fig:soap_spot_plages}
\end{figure*}

As we can tell from Fig.~\ref{fig:soap_spot_plages}, higher latitudes tend to have minor effects on $A_k$ ($\Delta A_k$ is shown using a solar CCF without spot or plage as a reference) and $\Delta RV_k$. This is because both the projected area of spots and plages in high latitudes and their rotational velocity along the line of sight are smaller. On the other hand, changes in $A_k$ and $\Delta RV_k$ per latitude towards the equator are not as prominent as in high latitudes, and this is because a latitude change near the equator does not change the the projected area of spots and plages along the line of sight and the rotation speed as much as in higher latitudes. 
% We do not further show the results for varying sizes as increasing spot/plage sizes behaves close to moving the spot/plage to lower latitudes. 

%-------------
%-------------

\begin{figure*}[t]
\centering
    \includegraphics[width=1\columnwidth]{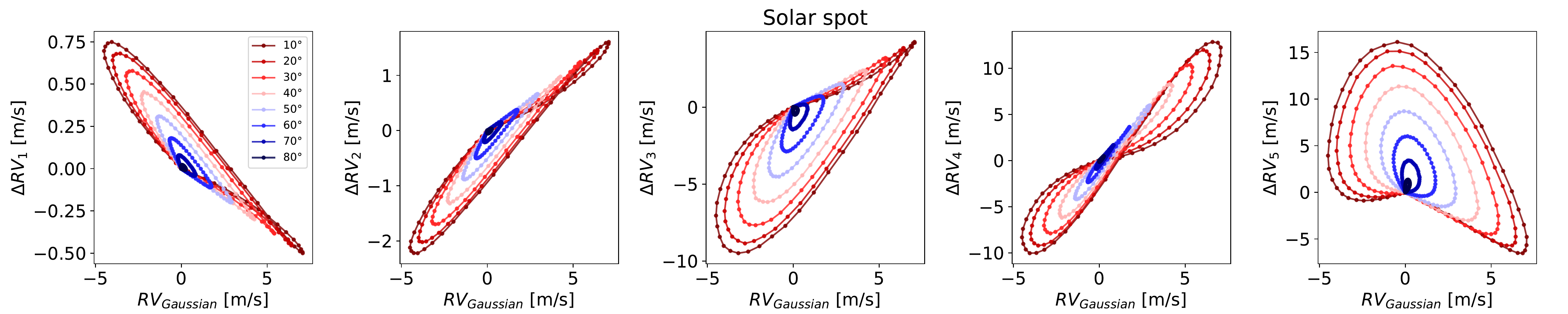}
    \includegraphics[width=1\columnwidth]{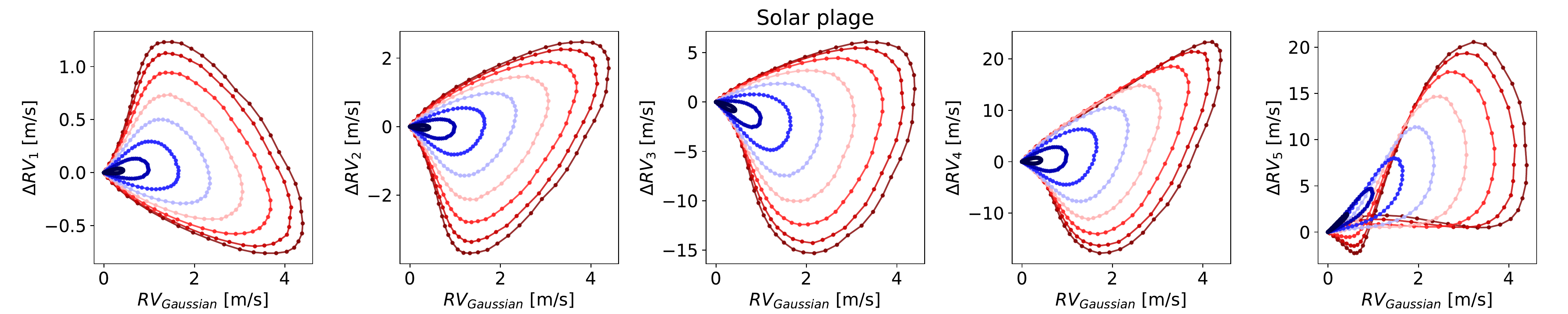}
    \caption{$\Delta RV_k$ and the apparent RV correlation for a simulated solar spot (top 5 panels) and a solar plage (bottom 5 panels) in one solar rotation. The latitudes are classified by the rainbow colours. Linear correlations between $\Delta RV_k$ and $RV_\text{Gaussian}$ are only prominent for spots in lower Fourier  modes ($k=1$ and 2). Spot and plage configurations are the same as the ones in Fig.~\ref{fig:soap_spot_plages}.
    % \ebf{Explain the input data set. What should readers take away from this figure?  }
    }
    \label{fig:soap_spot_plages-correlations}
\end{figure*}

Since changes in both $A_k$ and $\Delta RV_k$ are parametrisations for the amount of spectral line variability, a larger deformation, which is likely to lead to larger apparent radial velocity shifts, is expected to show larger changes in $A_k$ and larger $\Delta RV_k$ as well. However, this is not always the case as the correlations presented in Fig.~\ref{fig:soap_spot_plages-correlations}. Overall, larger apparent RV tend to be only associated with larger $\Delta RV_k$ for spots in lower Fourier  modes (e.g. $k=1$ and 2). For plages and higher Fourier  modes, their linear correlations are not observed.
% \ebf{Would it be fair to say that the spurious RV signal is concentrated in k=1,2, but the higher k's show a stronger dependence on latitude, so they may be valuable or even critical for figuring out what the cause of a spurious RV signal is?  Or is the SNR in k=3 so much smaller than k=1 and 2, then it's not more helpful for inferring the latitude?}
On the flip side, the different patterns of $\Delta RV_k$ - $RV_{\text{Gaussian}}$ plots and their stronger dependence on latitude will be valuable for figuring out the latitude information of the spots and plages. 
%\jzhao{add error bars in figure}
% \begin{figure}[t]
%     \begin{subfigure}{.49\textwidth}
%         \centering
%         \includegraphics[width=0.99\columnwidth]{figure/FIESTA-spot_latitude_correlation.png}
%     \end{subfigure}\hfill
%     \begin{subfigure}{.49\textwidth}
%         \centering
%         \includegraphics[width=0.99\columnwidth]{figure/FIESTA-plage_latitude_correlation.png}
%     \end{subfigure}
%     \caption{$\Delta RV_k$ and the apparent RV correlation in one solar rotation. The latitudes are classified by the rainbow colours and the modes $k$ are classified by the transparency of the dots, with the $k=1$ the most opaque and $k=3$ the most transparent and $k=2$ in between.
%     \ef{Explain the input data set. What should readers take away from this figure?  Should this figure be after the next one with simple timeseries?  }
%     \label{fig:soap_spot_plages-correlations}
% \end{figure}

%-------------
%-------------

%---------------------------------------------------
\subsection{Simulated continuous solar observations}
\label{sec:Simulated continuous solar  observations}
%---------------------------------------------------

In the next step, we analysed SOAP~2.0 solar simulations based on the known spot properties considering a similar activity level to the Sun, including spot temperature, size distribution and lifespan \citep{Gilbertson_2020}. The simulated data set was sampled twice per day and has 730 spectra covering a whole year. It was originally used for timeseries analysis \citep{Gilbertson_2020_II}, we nevertheless use the simulated solar CCFs as individual observations and seek correlations between $\Delta RV_k$ and the apparent RV due to solar spots.

\begin{figure*}[t]
\centering
    \includegraphics[width=0.9\columnwidth]{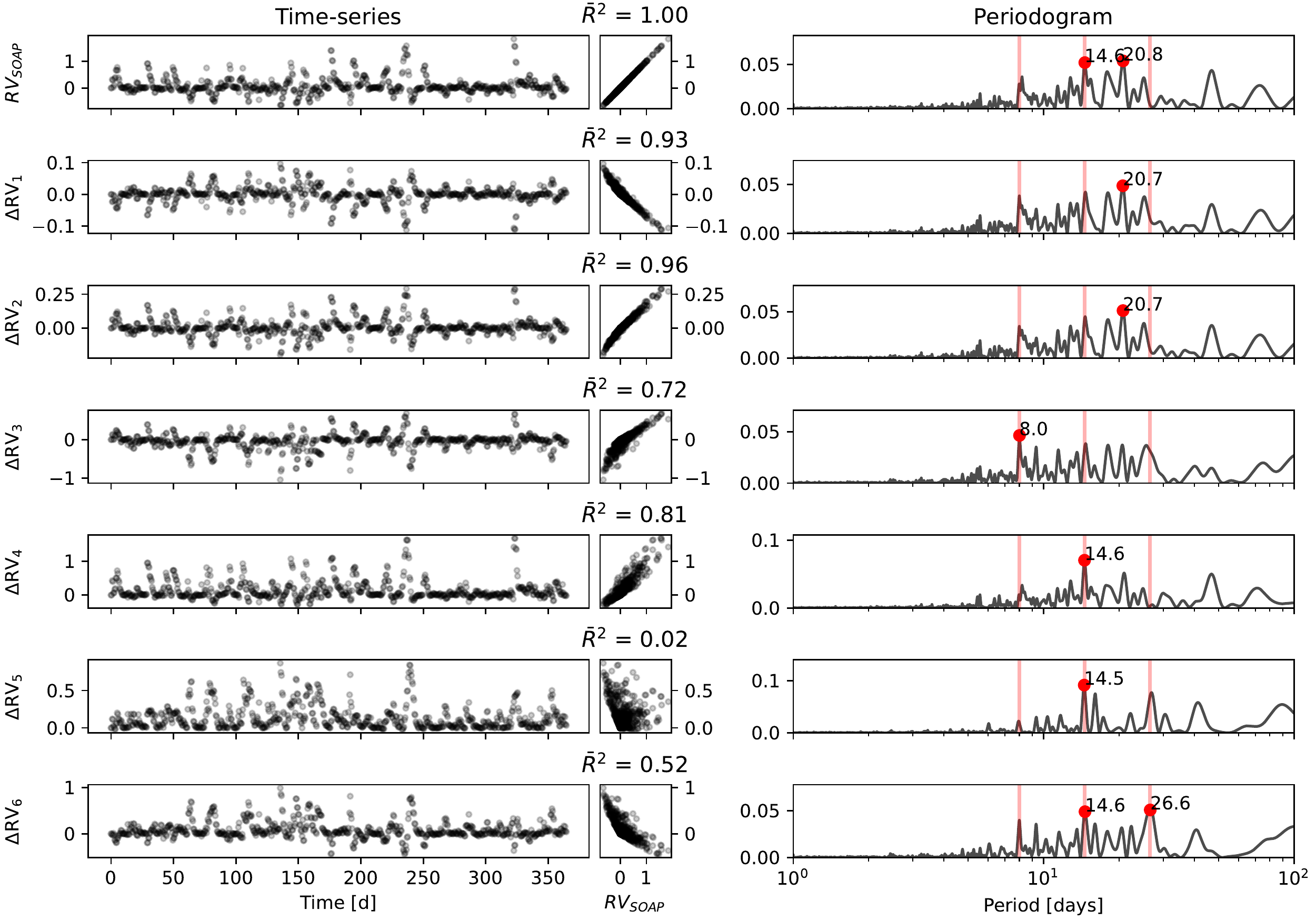}
    \caption{RV and $\Delta RV_k$ timeseries of SOAP~2.0 simulations of the Sun with evolving spots over a year. Units are in m/s. The lower Fourier  modes ($k=1$ and 2) show strong linear correlations with $RV_\text{SOAP}$, as quantified by the adjusted coefficient of determination $\bar{R}^2$ (Appendix~\ref{sec:coefficient_of_determination}). The corresponding periodograms are on the right, with red dots labelling the prominent periodicities and vertical lines labelling periodicities near the solar rotation and its harmonics at 26.6, 14.6 and 8.0 days.}
\label{fig:timeseries_soap}
\end{figure*}

We present the RV and $\Delta RV_k$ timeseries of the simulated data set in Fig.~\ref{fig:timeseries_soap}. The results are consistent with the discussions in Section~\ref{sec:spot_plage_correlations} that $\Delta RV_k$ shows strong linear correlations with the apparent RV ($RV_\text{SOAP}$) for $k=1$ and $2$, as the \citep{Gilbertson_2020_II} solar simulations are deformed only by spots but not plages. It indicates the \FIESTA lower Fourier modes may be used for linear RV correction for spot-dominated stars.
%-------------
%-------------

%%%%%%%%%%%%%%%%%%%%%%%%%%%%%%%%%%%%%%%%%%%%%%%%%%%%%
\section{\FIESTAtitle on HARPS-N 3-years solar observations}
\label{sec:FIESTA_HARPS}
%%%%%%%%%%%%%%%%%%%%%%%%%%%%%%%%%%%%%%%%%%%%%%%%%%%%%
%---------------------------------------------------
\subsection{Data}
\label{sec:data}
%---------------------------------------------------
As an example to show how \FIESTA can analyse real world line profile variability, we apply \FIESTA to the high-cadence HARPS-N spectroscopic observations of the Sun-as-a-star during 2015 to 2018 \citep{Dumusque_2015, Phillips_2016}. The data are publicly accessible on \url{https://dace.unige.ch/sun/}, including 34550 observations. These public data sets were already pre-selected to minimise non-astrophysical effects on observed CCF \citep{Cameron2019}.  They were taken in good observing condition (i.e. not significantly affected by clouds), the differential extinction for each observation less than 0.1~m/s, and the data reduction software (DRS) reports good quality results. The same selection criteria were also used in \cite{debeurs2020identifying} and \cite{Dumusque2021HARPS-N}. The HARPS-N CCFs and RVs used in \cite{Dumusque2021HARPS-N}, \cite{Cameron_2021} and in this paper (daily binned as in first row of Fig.~\ref{fig:time-series_and_shift_correlation}) were processed by the new HARPS-N DRS as opposed to those used in \cite{debeurs2020identifying} processed by an older version of the HARPS-N DRS. As a result, the $RV_\textrm{HARPS}$ presented in this paper reveals a long-term RV drift induced by the solar magnetic cycle
and contributes to a larger scatter of apparent RVs (both before and after applying mitigation procedures) than the ones in \cite{debeurs2020identifying}.

We work on the inverted, normalised daily weighted averaged CCF and the the daily weighted RVs, therefore reducing the number of original observations over the three years from 34550 to 567. Each exposure is designed to integrate over the solar p-mode oscillation timescale of $\sim$ 5 min \citep{Strassmeier2018, Chaplin2019}. The daily weighted average CCFs further average out both the p-mode solar oscillation effects and much of solar granulation which is significant on timescales from minutes for granules \citep{Ramio2005} to hours for mesogranules \citep{Matloch2009}. A \FIESTA study on these intra-day oscillation and granulation effects will be carried out in the near future. 
% %-------------
% \begin{figure}[t]
%     \includegraphics[width=0.49\columnwidth]{figure/rv_daily.png}
%     \includegraphics[width=0.49\columnwidth]{figure/rv_Periodogram.png}
%     \caption{Top: the unbinned and daily binned heliocentric RV timeseries of the three years of HARPS-N high-resolution spectroscopic data for the Sun. Bottom: the periodogram of the daily binned RVs with prominent periods labelled in orange.}
% \label{fig:harps-N-rv_daily}
% \end{figure}
% %-------------

The periodogram of the daily weighted HARPS-N RVs (first row of Fig.~\ref{fig:time-series_and_shift_correlation} or \ref{fig:time-series_and_A_correlation}) shows several prominent periodicities in the daily RV timeseries, with $\sim$28.5 days being identified as the solar rotation period. The other periods will be discussed in the following sections. The unbinned RV periodogram (not shown) is almost identical to the daily binned RV periodogram above 1 day.

%---------------------------------------------------
\subsection{\FIESTAtitle timeseries}
\label{sec:FIESTA_ts}
\begin{figure*}[t]
\centering
    \includegraphics[width=1\columnwidth]{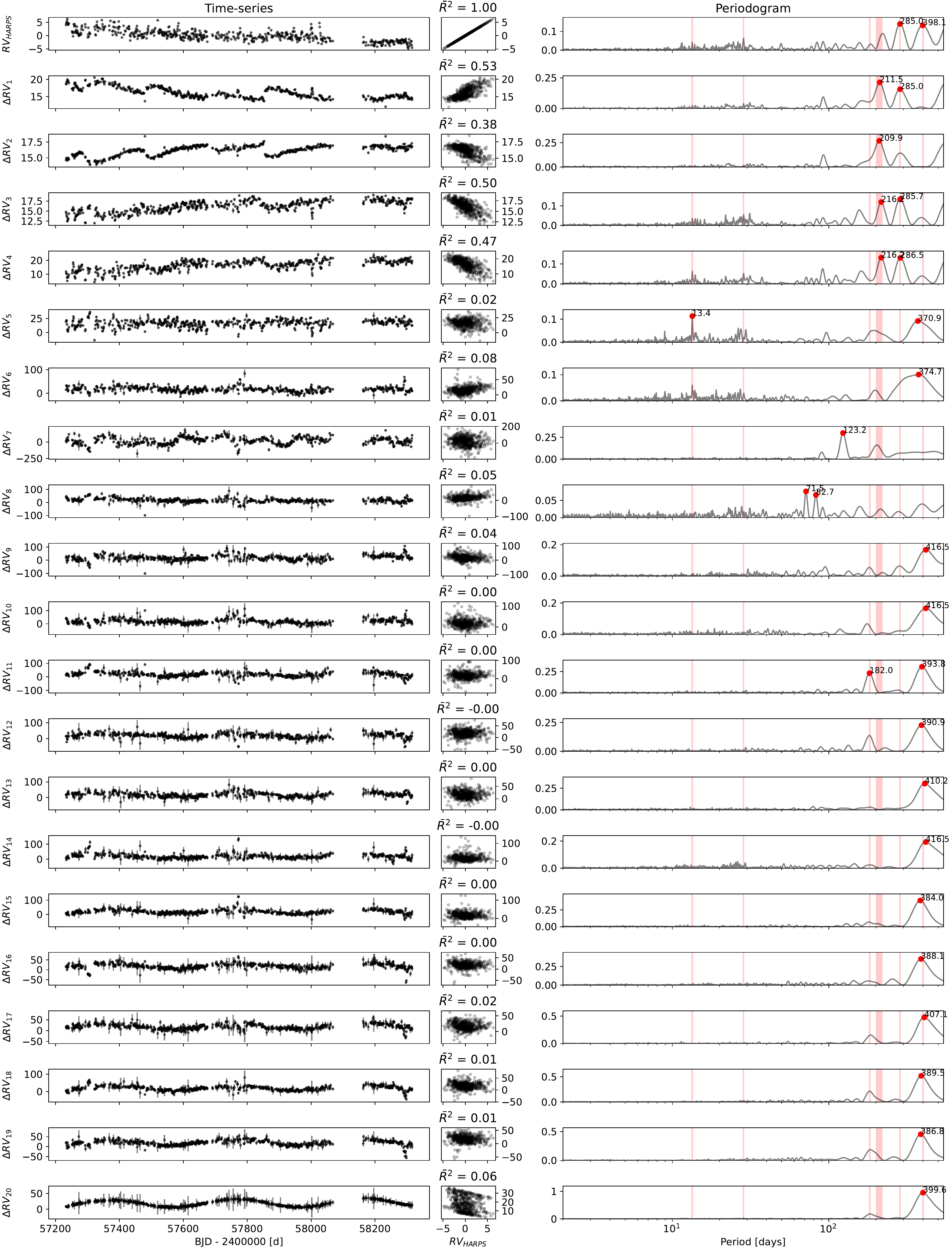}    
    \caption{Left: RV inferred from HARPS-N solar observations (top row) and $\Delta RV_k$ for the first 20 Fourier modes.
    $RV_\textrm{HARPS}$ and $\Delta RV_k$ timeseries have units of m/s. 
    Middle: The correlations between each $\Delta RV_k$ and $RV_\textrm{HARPS}$ described by the adjusted coefficient of determination $\bar{R}^2$.
    Right: The periodograms of the corresponding timeseries with prominent periods labelled in red dots. The vertical bands correspond to periods at 13.4, 28.5, 365/2, 200-220, 285 and 400 days as discussed in Section~\ref{sec:FIESTA_ts}.}
    \label{fig:time-series_and_shift_correlation}
\end{figure*}

\begin{figure*}[t]
\centering
    \includegraphics[width=1\columnwidth]{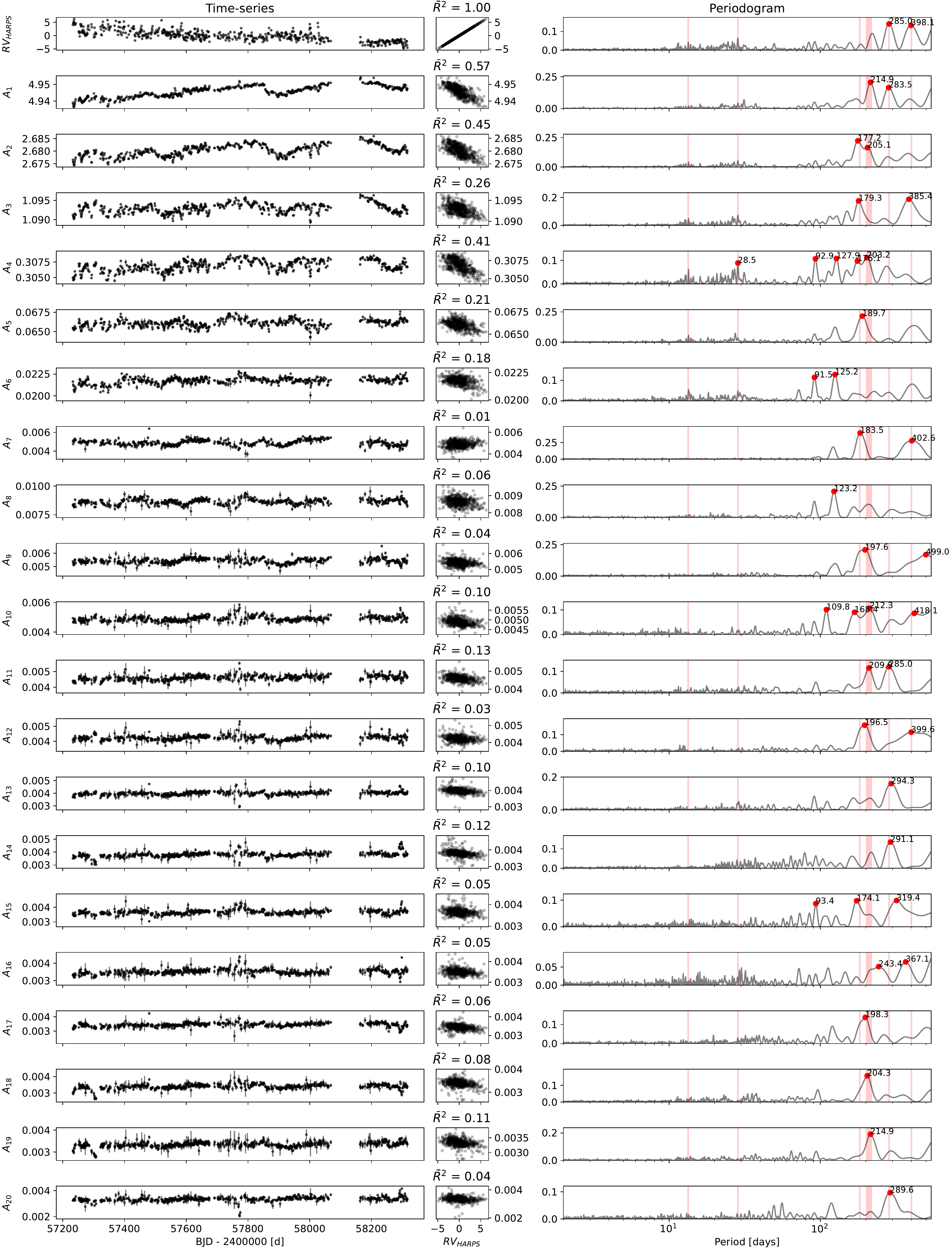}
    \caption{Same as Fig.~\ref{fig:time-series_and_shift_correlation}, but for $A_k$ instead of $\Delta RV_k$.}
    \label{fig:time-series_and_A_correlation}
\end{figure*}

%---------------------------------------------------
In the left-hand panels of Fig.~\ref{fig:time-series_and_shift_correlation} and \ref{fig:time-series_and_A_correlation}, we show the timeseries for each $\Delta RV_k$ and $A_k$ for 20 Fourier modes. 
Different modes characterise the CCF deformation on different scales, with $k=1$ being the longest mode (refer to Fig.~\ref{fig:Flux_residual}). 
As $k$ increases, the scale of study decreases as $1/k$ and the amplitude decays approximately as a Gaussian function of $k$ (Fig.~\ref{fig:challenging test}). Therefore, the signal-to-noise ratio for  measurements of $\Delta RV_k$ and $A_k$ become dominated by photon noise for large $k$. 
The middle panels are the correlations between each row's indicator and $RV_\text{HARPS}$.
The right-hand panels of Fig.~\ref{fig:time-series_and_shift_correlation} and \ref{fig:time-series_and_A_correlation} show the periodogram for each $\Delta RV_k$ and $A_k$.

Based on the $\Delta RV_k$ timeseries and the timescales observed in their periodograms, we've identified five mechanisms contributing to the variations in the CCF profile:
\begin{enumerate}
    \item Solar rotationally modulated CCF variations: The $\sim28$ day solar rotation period and its harmonics can be seen in both $\Delta RV_k$ and $A_k$ periodograms (e.g. $k$ from 3 to 8). The relative significance of the peaks often differs between the periodograms of $\Delta RV_k$ and $A_k$, even for the same $k$.  Rotationally modulated variations are more obvious in $\Delta RV_k$ periodograms.
    \item Solar magnetic cycle: The long-term RV drift over the three years of observations is correlated with $\Delta RV_k$ and $A_k$ up to $k=4$ ($\bar{R}^2\sim0.5$).
    \item CCD detector warm-ups: Instrumental features can be seen most clearly as the sudden jumps of $\Delta RV_1$ and $\Delta RV_2$ that coincide with times of interventions to the CCD detector \citep{Dumusque2021HARPS-N} . While not strictly periodic, these this effect likely leads to the periodogram peaks near $\sim 210$~d and $\sim 285$~d.
    \item Changes in the apparent solar angular velocity: The Earth-Sun distance changes both due to Earth's eccentricity and as a result of perturbations from Jupiter \citep[also discussed in][]{Cameron2019}. The periodicity at around 400 days is predominant in the higher Fourier modes (e.g. $k\geqslant9$) and nearly matches the time between conjunctions with Jupiter.
    \item Changes in the apparent solar angular velocity as a result of the Earth's orbit being slightly eccentric changing the distance between Earth and the Sun \citep{Cameron2019}. This periodicity at half a year is apparent in many of the higher Fourier modes, but can be overwhelmed by the CCD detector warm-ups and/or harmonics of the timescale between conjunctions of Jupiter. 
\end{enumerate}
%-------------
%-------------
%-------------
%-------------

While both $\Delta RV_k$ and the change in $A_k$ trace the CCF variability, $A_k$ seems to capture more periodicities. This is expected since line depth changes or artifacts in the estimate for the continuum around the line may affect $A_k$'s, but not $\Delta RV_k$'s.
% We note that changes in the $A_k$'s could be due to either the actual line depth changing or changes in the estimate for the continuum around the line.
% The fact that $\Delta RV_k$ periodograms are cleaner than the $A_k$ periodograms can be seen as a form of verification that \FIESTA's $\Delta RV_k$'s are contributing to the long-term goal of robustly distinguishing stellar variability from instrumental and telluric effects. 

%---------------------------------------------------
\subsection{Principal component analysis}
\label{sec:PCA}
%---------------------------------------------------
Some of the nearby Fourier modes present similar or correlated behaviour as seen in the timeseries and their periodograms, especially for $\Delta RV_1$ and $\Delta RV_2$, which appear anti-correlated with each other. It is possible that certain features of a line deformation can take place across several Fourier  modes and thus cause correlated effects. Therefore, we can model a large fraction of the CCF variance with a smaller number of coefficients by invoking principal component analysis (PCA) to perform dimensional reduction on the $\Delta RV_k$'s (or $A_k$'s).  This has dual benefits of increasing the SNR in the retained PC scores and potentially ease interpretation of a lower dimensional timeseries.

In an ideal world with ultra-high SNR, we could make use of all the 20 Fourier modes for the CCF parametrisation. In reality, higher Fourier modes are more strongly affected by measurement uncertainties; in addition, they do not contribute as much information due to rapidly decaying amplitudes (Section~\ref{sec:Measurement_uncertainties}). As mentioned in Section~\ref{sec:Summary_notes}, it is advised to use no more than 11 Fourier modes for the HARPS-N solar data analysis. If we were to focus on the first 3 principal components built from the first $k_\textrm{max}$ modes, we find that increasing $k_\textrm{max}$ from 7 to 11 hardly changes the first 2 PC score timeseries, and increasing $k_\textrm{max}$ from 9 to 11 hardly changes the $3^{\textrm{rd}}$ PC score timeseries.  Therefore, we choose $k_\textrm{max}=9$ to parametrise the CCF variability.  When performing the analysis in Sections~\ref{sec:mlr} and \ref{sec:mlr_with_lags}, we also find that building the first 3 principal components from the first 9 Fourier modes results in the smallest residual (compared with other $k_\textrm{max}$), though the difference of the residual WRMS (i.e. weighted RMS) is less than 5\% among choices of $k_\textrm{max}$ ranging from 7 to 11.

\begin{figure*}[t]
\centering
    \includegraphics[width=0.99\columnwidth]{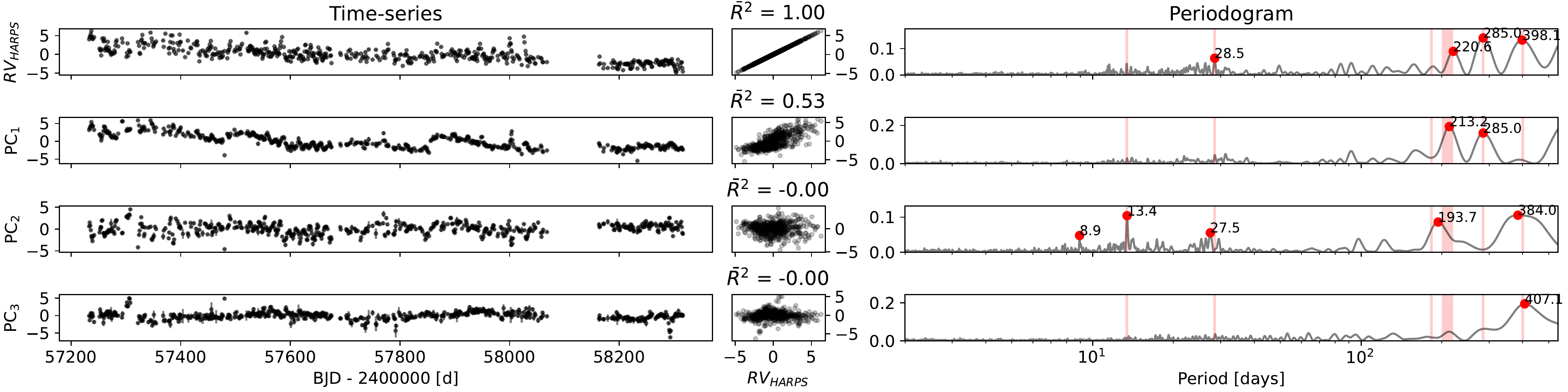}
    \caption{Left: PCA for $\Delta RV_k$ up to $k_\textrm{max}=9$. The first 3 principal components explain 75\% of the variance in $\Delta RV_k$ $(k=1,2,\dots,9)$. Middle and the right columns are the same correlations plots and periodograms plots as in Fig.~\ref{fig:time-series_and_shift_correlation} except for these principal components.}
    \label{fig:FIESTA_RV_PCA}
\end{figure*}

While most PCA packages are designed to deal with equally weighted data by default, the HARPS-N observations have associated errorbars. The weighted PCA is aimed to tackle this problem by solving the weighted covariance matrix of the data \citep{Delchambre2015}. We used the \texttt{wpca} python package for its implementation. The weighted PCA on $\Delta RV_k$ (Fig.~\ref{fig:FIESTA_RV_PCA}) reduces the feature dimension from 9 to 3 while $\sim$75\% of variance in the data are preserved. The residual due to projecting data into a lower dimension accounts for the rest $\sim$25\% of the variance. We can tell the aforementioned periodicities in Section~\ref{sec:FIESTA_ts} are preserved in the first 3 PC scores. Note that we normalise the $\Delta RV_k$ for each $k$ before applying weighted PCA to prevent the higher modes from dominating the principal components.

%-------------
%-------------
% %-------------
% \begin{figure*}[t]
% \centering
%     \includegraphics[width=0.99\columnwidth]{figure/PCA_A_k_max=7.png}
%     \caption{Same as Fig.~\ref{fig:FIESTA_RV_PCA} but for $A_k$ up to $k=7$.}
%     \label{fig:FIESTA_amplitude_PCA}
% \end{figure*}
% %-------------
% We show the results of applying PCA to the first $N_k=7$ $A_k$'s  (Fig.~\ref{fig:FIESTA_amplitude_PCA}) are for demonstration only as we do not use amplitude information directly for the CCF paramtrisation.

% Interestingly, the solar rotation period signal $\sim$28.5 days in the $A_k$ periodogram (Fig.~\ref{fig:Periodogram_fiesta_harps}) disappeared in the first 6 principal components (Fig.~\ref{fig:FIESTA_amplitude_PCA}). However, the rotation period and its harmonic present in the $\Delta RV_k$ periodogram (Fig.~\ref{fig:Periodogram_fiesta_harps}) becomes a prominent feature at 13.4 days in the second principal component (Fig.~\ref{fig:FIESTA_RV_PCA}). Both $A_k$ and $\Delta RV_k$ tend to show a periodicity around half a year (dashed line in both PC scores periodograms), while the other long term feature around 400 days are present in multiple PC scores periodograms as well as the original ones. 

%\ebf{First, you should check the dates mentioned in the Collier-Cameron, Ford et al 2021 paper where we found CCF shape chagnes associated with some detector intervention (something about a cold plate).  Also it would be nice to try a wPCA on joint Ak's and RVk's.} 

%---------------------------------------------------
\subsection{Separating out long-term and short-term variations}
\label{sec:ls-variations}
\begin{figure*}[t]
\centering
    \includegraphics[width=0.99\columnwidth]{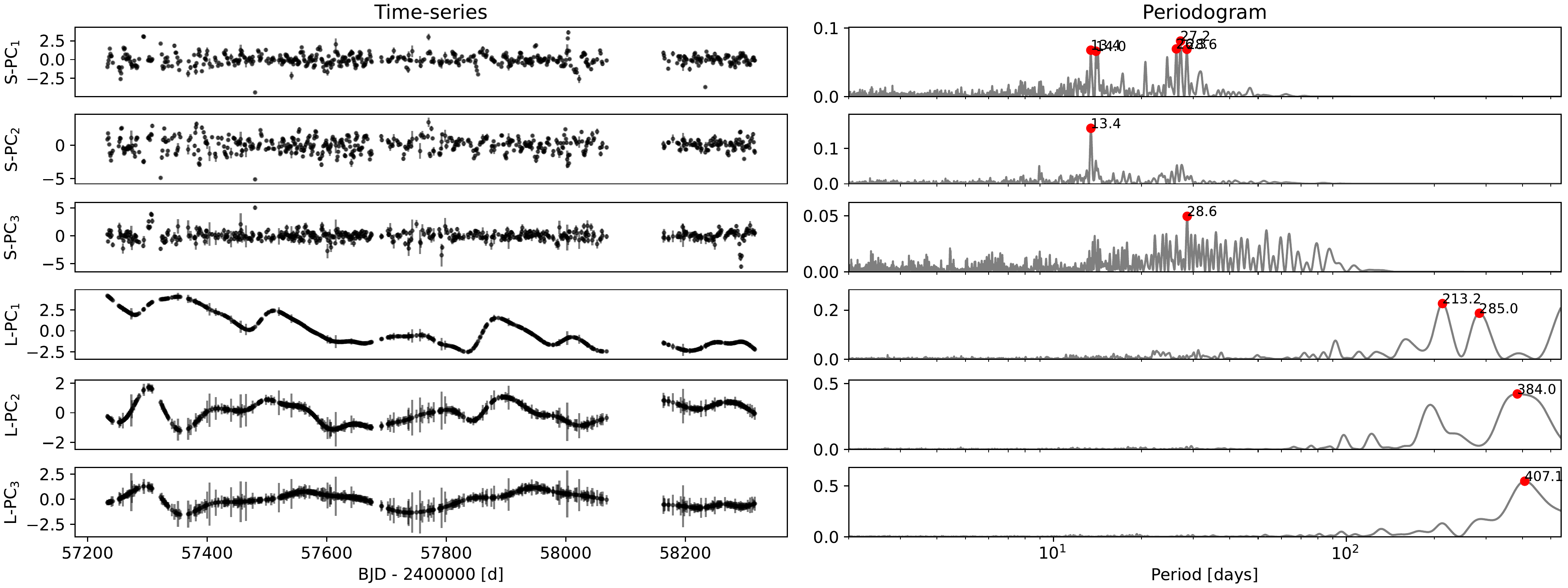}
    \caption{Separating the short-term (S-PC) and long-term (L-PC) variability of the 3 principal components in Fig.~\ref{fig:FIESTA_RV_PCA}. The corresponding periodograms are on the right.}
    \label{fig:ls-variation}
\end{figure*}

%---------------------------------------------------
PCA extracts independent features in high dimensional data while keeping most of the variation information within the first few principal components, however, the process is not physically driven and thus we should not expected that each mechanism that contributes to CCF variations will be neatly contained in an individual principal component. Therefore, we apply a low-pass filter to the PC scores to separate rotationally induced effects (periods $<100$ days) from long-term effects (periods $>100$ days).
Specifically, we compute L-PC$_i(t)$, the mean of Gaussian process conditioned on each of the PC score timeseries (PC$_i$'s) separately using a Mat\'ern 5/2 kernel (\citealp[e.g.][]{Genton2001, Matern2005}) with a length-scale of $r=100$ days, which captures the variations above $r$ and smooths out those below $r$ (Fig.~\ref{fig:ls-variation}). We then subtract L-PC$_i$ from the corresponding principal component scores to obtain S-PC$_i$, which probes the short-term variability below $\sim$100 days and are found to be dominated by the solar rotationally modulation.
%-------------
%-------------

%---------------------------------------------------
\subsection{Multiple linear regression modelling}
\label{sec:mlr}
%---------------------------------------------------
Once PCA has reduced the number of features required to accurately parametrise the CCF variation, we want to investigate whether these PC scores are useful for predicting the behaviour of the spurious RV measurements. The simplest approach would be a multiple linear regression model
% matrix form:
\begin{equation}
    y \equiv RV_\text{HARPS} = \mathbf{X}\beta + \epsilon
\end{equation}
where $\mathbf{X}$ is the regressor matrix that traces the CCF variability, $\beta$ is the coefficient vector, $\epsilon$ is the error vector and $y \equiv RV_\text{HARPS}$ is the response vector. 

We examine two models for which $\mathbf{X}$ consists of various features derived from the CCFs parametrisation, Model 1 using the first three principal component scores (PC$_i$'s computed from $\Delta RV_k$, $k=1,2...,9$, Fig.~\ref{fig:FIESTA_RV_PCA}) and Model 2 taking a step further by using their short-term components $\textrm{S-PC}_i (i=1,2,3)$ and long-term components $\textrm{L-PC}_i (i=1,2,3)$ as regressors (Fig.~\ref{fig:ls-variation}). As models naturally fit the data better with more features or parameters, to avoid over-fitting, we perform Lasso regression (i.e., added a L1 regularisation penalty to the weighted sum of the squared residuals, \citealp{Tibshirani_lasso, alma991004158939705164}) and minimise the following
%The factor $\lambda$ specifies how strong of a regularisation penaly is applied and controls the number of non-zero coefficients: 
%
\begin{equation}
    \text{loss function} = (y - \mathbf{X}\beta)^T \mathbf{W} (y - \mathbf{X}\beta) + \lambda \|\beta\|_1
\label{eq:lasso}
\end{equation}
where $\mathbf{W}$ is the weight matrix, $\lambda$ is the tuning parameter and $\|\beta\|_1$ is the 1-norm of the vector $\beta$ defined as $\sum_{i=1}^{n} |\beta_i|$. 
$\lambda=0$ implies no penalty and Eq.~\ref{eq:lasso} is reduced to the normal linear regression optimisation.
The larger $\lambda$ becomes, the larger the penalty term and the more coefficients are driven to zero. 
To find out the optimal $\lambda$, we adopt a cross-validation approach where the data are randomly scrambled and split into 5 folds. We train our model as we minimise the loss function in Eq.~\ref{eq:lasso} using 4 folds of the RV data and feature matrix, and then obtain a residual WRMS for the validation data. This process is repeated 100 times, each with a randomised split and $\lambda$ is tested from $10^{-3}$ to 1. Then we locate $\lambda=\hat{\lambda}_1$ corresponding to the smallest WRMS in the testing set. Usually a more regularised and simpler model with $\lambda>\hat{\lambda}_1$ is preferred.

With the multiple linear regression model regularised by the loss function above (Eq.~\ref{eq:lasso}), we reduce the uncorrected WRMS of $RV_\textrm{HARPS}$ from 2.00~m/s to 1.36~m/s for using the first 3 PC scores (Fig.~\ref{fig:FIESTA_RV_PCA}) and further down to 1.14~m/s with additionally separating out the long-term and short-term variations (Fig.~\ref{fig:ls-variation}), corresponding to 32\% and 43\% reduction in the WRMS, or 53\% and 67\% reduction in the weighted variance, as calculated by $\bar{R}^2$. Among the total reduction in the weighted variance, we determine the importance of each feature by calculating their contributed variance $(\beta_i\sigma_i)^2$. The information is summarised in Table~\ref{table:mlr1}'s rows corresponding to Models 1 and 2. 

We estimate that solar rotation-induced RVs contributes just $\sim7\%$ of the variance (based on sum of the variance percentages due to $\textrm{S-PC}_i (i=1,2,3)$) or $\sim0.4$~m/s in terms of RMS without considering measurement errors or instrumental instability.
In contrast, $\textrm{L-PC}_1$ (which traces the solar magnetic cycle and the CCD detector warm-ups) contributes over 80\% of the variance reduction. Most of the remaining $\sim$10\% of the RV variation is believed to be caused by changes in the apparent solar rotation rate due to Jupiter and Earth's eccentricity ($\textrm{L-PC}_2$).

\begin{table}
\centering
    % New table. PCA from 9 Fourier modes 
    \begin{tabular}{*{10}c}
    Model               &Label              &$\beta_i$  &$\sigma_i$         &$(\beta_i\sigma_i)^2$     &Data WRMS  &Model WRMS   &Residual WRMS     &$\bar{R}^2$\\
    \hline
    
    \multirow{3}{*}{1}  &$\textrm{PC}_1$&   0.73&       1.96&               2.06 (99.9\%)&              \multirow{3}{*}{2.00}& \multirow{3}{*}{1.44}& \multirow{3}{*}{$1.36~(-32\%)$}&    \multirow{3}{*}{0.532}\\
                        &$\textrm{PC}_2$&   -0.00&      1.25&               0.00 (0.0\%)\\
                        &$\textrm{PC}_3$&   0.04&       1.06&               0.00 (0.1\%)\\  
    \hline
    
    \multirow{6}{*}{2}  &$\textrm{S-PC}_1$  &0.51       &0.71               &0.13 (4.8\%)               &\multirow{6}{*}{2.00}& \multirow{6}{*}{1.55}& \multirow{6}{*}{$1.14~(-43\%)$}     &\multirow{6}{*}{0.669}\\
                        &$\textrm{S-PC}_2$  &0.23       &1.05               &0.06 (2.1\%)\\
                        &$\textrm{S-PC}_3$  &-0.03      &0.81               &0.00 (0.0\%)\\
                        &$\textrm{L-PC}_1$  &0.83       &1.81               &2.24 (80.7\%)\\
                        &$\textrm{L-PC}_2$  &-0.87      &0.63               &0.30 (10.8\%)\\
                        &$\textrm{L-PC}_3$  &0.35       &0.62               &0.05 (1.7\%)\\
    \hline                      
    
    3                   &\multicolumn{4}{c}{(see Fig.~\ref{fig:lasso_var})}                             &1.89 &1.53                   &$0.94~(-50\%)$                     &0.733\\
    \hline 

    \multirow{2}{*}{4}& FWHM&               0.29&       0.95&               0.08 (4.0\%)&               \multirow{2}{*}{2.00}& \multirow{2}{*}{1.43}& \multirow{2}{*}{$1.34~(-33\%)$}&    \multirow{2}{*}{0.552}\\
                      & BIS&                1.35&       1.00&               1.79 (96.0\%)\\
    \hline

    \multirow{4}{*}{5}& S-FWHM&             0.70&       0.70&               0.24 (10.4\%)&              \multirow{4}{*}{2.00}& \multirow{4}{*}{1.54}& \multirow{4}{*}{$1.18~(-41\%)$}&    \multirow{4}{*}{0.651}\\
                      & S-BIS&              0.00&       0.48&               0.00 (0.0\%)\\
                      & L-FWHM&             0.16&       0.59&               0.01 (0.4\%)\\
                      & L-BIS&              1.66&       0.86&               2.06 (89.2\%)\\                      
    \hline
    
    6                 & \multicolumn{4}{c}{(see Fig.~\ref{fig:lasso_fb})}                               &1.89 &1.46                   &$1.04~(-45\%)$                     &0.680\\
    \hline     
    \end{tabular}
    
    \caption{Comparisons of 6 models fitting the HARPS-N solar spurious RV. The difference among the models are the regressors used in the multiple linear regression model. Note that the adjusted coefficient of determination $\bar{R}^2$ accounts for the the number of free parameters in a model and so $\bar{R}^2$ can indicate the performance of a model regardless of different number of parameters (Appendix~\ref{sec:coefficient_of_determination}). \\
    Model 1: using PC scores derived from $\Delta RV_k$ ($k_\textrm{max} = 9$). \\
    Model 2: in additional to Model 1, using the solar rotation components $\textrm{S-PC}_i$ and the long-term variation components $\textrm{L-PC}_i$ derived from the PC scores are separated out. \\
    Model 3: in additional to Model 2, introducing time lags for the solar rotation components $\textrm{S-PC}_i$. \\
    Models 4 - 6: repeating for Models 1-3 but for FWHM and BIS.}
    \label{table:mlr1}
\end{table}

%---------------------------------------------------
\subsection{Multiple linear regression modelling with time lags}
\label{sec:mlr_with_lags}
%---------------------------------------------------
\cite{Cameron2019} explored the correlation between the spurious RV signal in HARPS-N observations and two traditional CCF shape indicators -- the Bisector Inverse Slope (BIS) and the full-width half-maximum (FWHM).
They found that the apparent RV variations led the BIS by 3 days and the FWHM by 1 day. We want to investigate if introducing a lag between $RV_\textrm{HARPS}$ and the principal components of $\Delta RV_k$ can improve our predictions for the spurious RV signal. As the lag between two timeseries can be poorly constrained when the variation timescale are too long compared to the time lag, we decide to fix a zero lag for the long-term variations $\textrm{L-PC}_i$ and only introduce a lag parameter in integer days for the short-term variations $\textrm{S-PC}_i$. 

In order to implement multiple linear regression with time lags, we must address two challenges. 
First, observations are not available for every consecutive day. Most gaps are less than a few days whilst a large one starts from 2017-11-11 (BJD$-2400000 = 58068.5$) due to the HARPS-N Fabry-P\'erot upgrade \citep{Dumusque2021HARPS-N}. 
Second, the daily binned timestamps are not exactly evenly spaced, creating a slight misalignment between timeseries when they are shifted by integer number of days. 
To overcome these challenges, 
%in an effort to study the time lags between the apparent RV and the CCF variability proxy indicators,
we linearly interpolate the indicators timeseries to the $RV_\textrm{HARPS}$ timestamps and their offsets in days. The beginning and the end of the two trunks of $RV_\textrm{HARPS}$ separated by the Fabry-P\'erot upgrade gap are not used so that we can avoid the artifacts of extrapolating the indicators timeseries outside the $RV_\textrm{HARPS}$ timestamps when dealing with lags. We also note that over 90\% of the daily binned observations are within 2 hours of the day, so using the $RV_\textrm{HARPS}$ directly as the response vector is sufficient to study the time lags in days.

% The daily interpolation almost doubles the data size of $RV_\textrm{HARPS}$ from 469 before 2017-11-11 to 835 timestamps, since  nearly half of the days before 2017-11-11 had no observations. The WRMS for the 835 interpolated $RV_\textrm{HARPS}$ becomes 1.70~m/s. 

\begin{figure*}[t]
\centering
    \includegraphics[width=0.49\columnwidth]{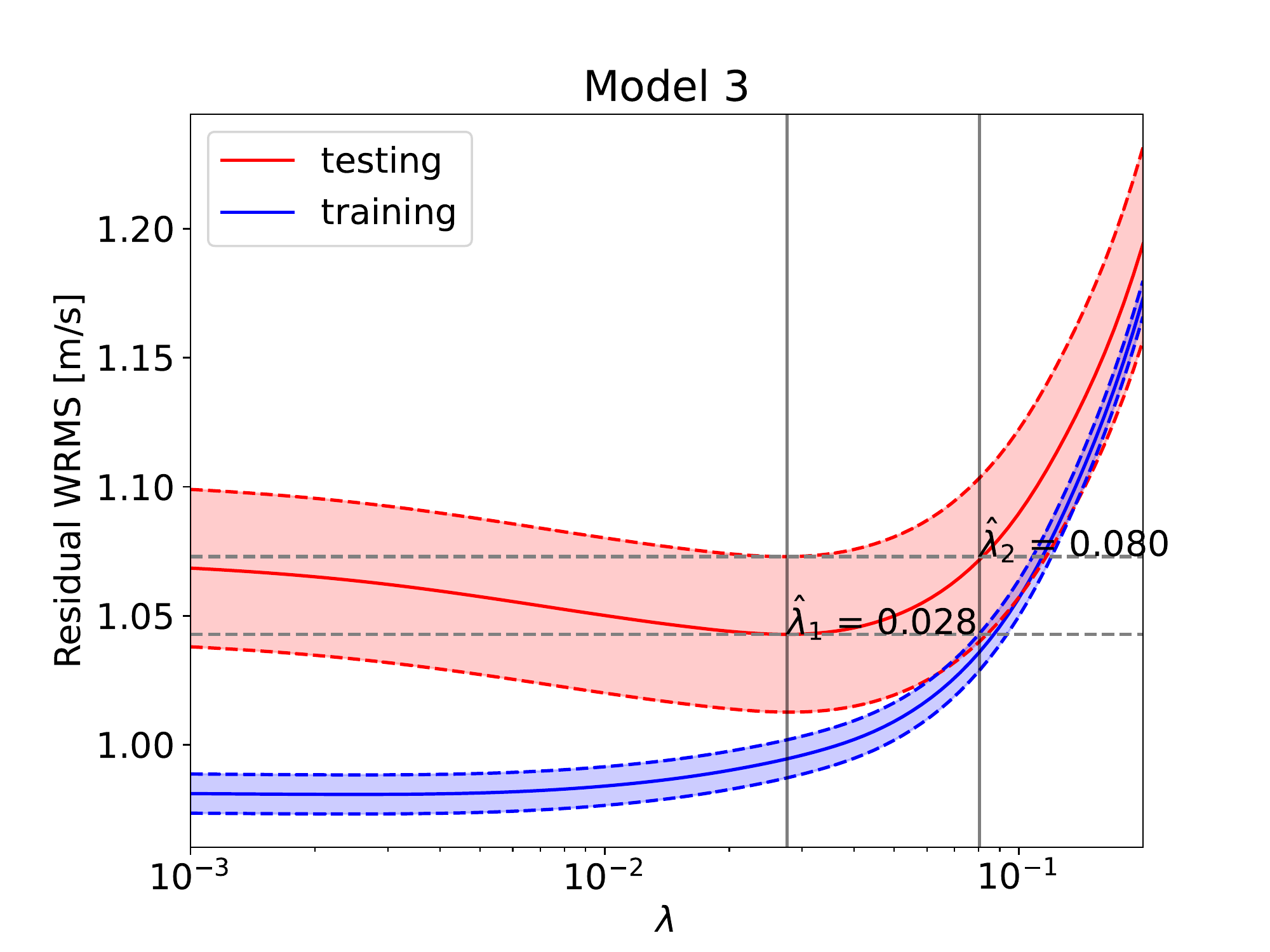}
    \includegraphics[width=0.49\columnwidth]{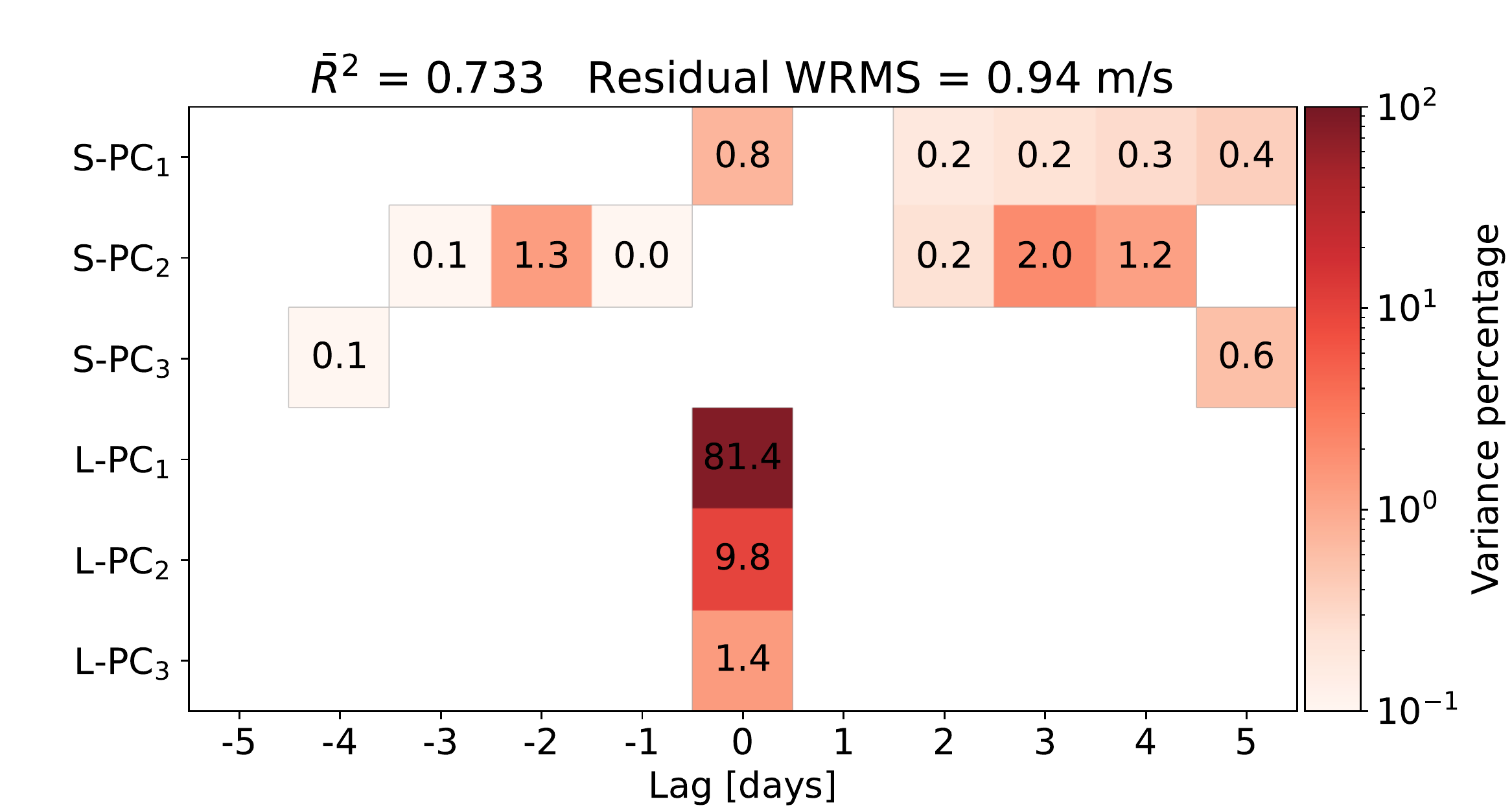}
    \caption{Left: we determine the tuning parameter $\lambda$ in the Lasso regression model (Eq.~\ref{eq:lasso}) as follows.
    Let $\lambda=\hat{\lambda}_1$ correspond to the smallest WRMS in the testing set (red solid line; with the red dashed indicating the $1\sigma$ uncertainty). Generally, a more regularised and simpler model with a larger $\lambda$ up to $\hat{\lambda}_2$ is preferred, where $\text{WRMS}(\hat{\lambda}_2) = \text{WRMS}(\hat{\lambda}_1) + \sigma_{\text{WRMS}(\hat{\lambda}_1)}$. We choose $\lambda=0.05$ in order to be consistent with other models.
    Right: the variance in percentage for each $\textrm{S-PC}_i$ and $\textrm{L-PC}_i$ ($i=1,2,3$) in the regularised multiple linear regression fitting of $RV_\textrm{HARPS}$. The blanks indicate zero coefficients (thus zero variance) due to the regularisation with $\lambda=0.05$, while the 0.0's are the result of rounding from a non-zero, but small enough variance. A positive lag means the proxy indicator lags behind the RV variation, i.e., the RV variation leads the proxy indicator. The model returns $\bar{R}^2=0.733$ and a residual WRMS of 0.94~m/s.}
    \label{fig:lasso_var}
\end{figure*}

The lagged multiple linear regression introduces additional fitting parameters -- one coefficient for each lag for each of the $\textrm{S-PC}_i$ and $\textrm{L-PC}_i$ ($i=1,2,3$).
We tested different maximum lags allowed and found that the fitted coefficients remain mostly stable once the model includes at least a maximum 5-day lag. Therefore, we present the results using a maximum 5-day lag. As for the regularisation parameter $\lambda$, we note the residual WRMS reaches minimum at $\hat{\lambda}_1=0.028$. According to ``the one standard error rule'' \citep{BreiFrieStonOlsh84, hastie_09}, a more regularised model with $\hat{\lambda}_2=0.080$ at one standard error of $\textrm{WRMS}(\hat{\lambda}_1)$ is usually recommended (Fig.~\ref{fig:lasso_var}). However, considering we would like to compare all the models using $\bar{R}^2$, it would be easier if the same $\lambda$ is chosen. Therefore, we choose the middle ground $\lambda=0.05$, which results in a more parsimonious Model 3 than if $\lambda=\hat{\lambda}_1$ were chosen and meanwhile $\lambda=0.05$ does not over regularise Models 1 and 2\footnote{For Models 1 and 2, their residual WRMS curves remain mostly flat until they start rising at $\lambda\sim0.01$, indicating hardly any overfitting and so regularisation is not needed. However, in order to fairly compare different models, we choose $\lambda=0.05$ for all the models.}. With this, we obtain a residual WRMS~=~0.94~m/s, a 50\% reduction in the WRMS or 73\% reduction in the weighted variance.
%-------------
%-------------

In Fig.~\ref{fig:lasso_var}, we observe that $\sim7\%$ of the variance (or 0.51~m/s in terms of RMS) comes from the solar rotationally induced RV variability (i.e., sum of the variance in $\textrm{S-PC}_i (i=1,2,3)$ for all lags), consistent with the $\sim7\%$ in Model 2 with no lags. Over 80\% of variability is dominated by terms that show trends of the solar magnetic cycle plus the CCD detector warm-ups. 
In addition, the lag information of $\textrm{S-PC}_i (i=1,2,3)$ tracing solar rotation is spread out between $-2$~days and 5~days, with the single most prominent lag for 3~days by summing up the variance of $\textrm{S-PC}_i (i=1,2,3)$ by day, accounting for $\sim2\%$ of the RV variance. 
Individually, the weighted averaged lag for $\textrm{S-PC}_i (i=1,2,3)$ are 2.2, 1.7 and 3.7~days, while \cite{Cameron2019} noted lags of 3 and 1 days between the BIS and FWHM and the spurious RV. 
Our finding that a weighted average of the indicators at multiple lags can provide a more effective predictor of the spurious RV signals is complementary to results from  \cite{Cameron2019}.
We hypothesise the different mechanisms that result in RV variation as the Sun rotates may show different lags, and $\textrm{S-PC}_i (i=1,2,3)$, BIS and FWHM detect part or a combination of these mechanisms. 

\section{The alternative: BIS and FWHM}
\label{sec:BIS-FWHM}
\begin{figure}[t]
    \includegraphics[width=1\columnwidth]{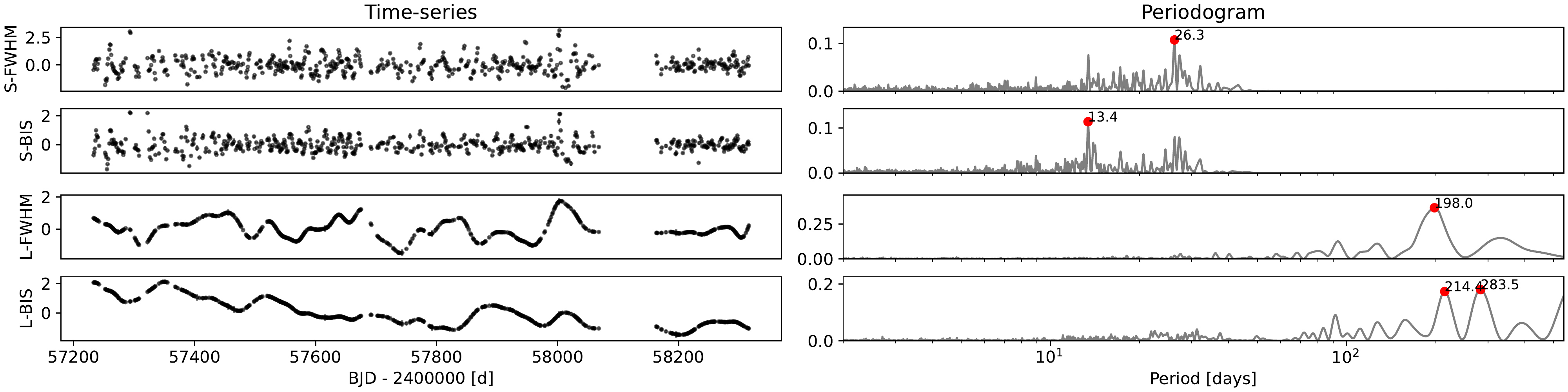}
    \caption{Same as Fig.~\ref{fig:ls-variation} but for FWHM and BIS.}
\label{fig:fwhm_bis}
\end{figure}

%---------------------------------------------------
To investigate if using \FIESTA metrics improves the modelling of line profile variability, we compare the results of \S \ref{sec:mlr} and \ref{sec:mlr_with_lags} with a similar analysis applied to two traditional CCF variability indicators, the FWHM and BIS. We repeat the process of HARPS-N RV modelling except for substituting the PC scores of $\Delta RV_k$ by the daily binned FWHM and BIS (Fig.~\ref{fig:fwhm_bis}). The originally unbinned FWHM and BIS data were provided on \url{https://dace.unige.ch/sun/}. The FWHM and BIS timeseries are standardised (mean = 0 and variance = 1) so that the linear regression coefficients are not penalised inappropriately in the Lasso regression because of their different units or because of one having intrinsically smaller variance but requiring a larger coefficient to compensate. 

%-------------
%-------------

\begin{figure*}[t]
\centering
    \includegraphics[width=0.49\columnwidth]{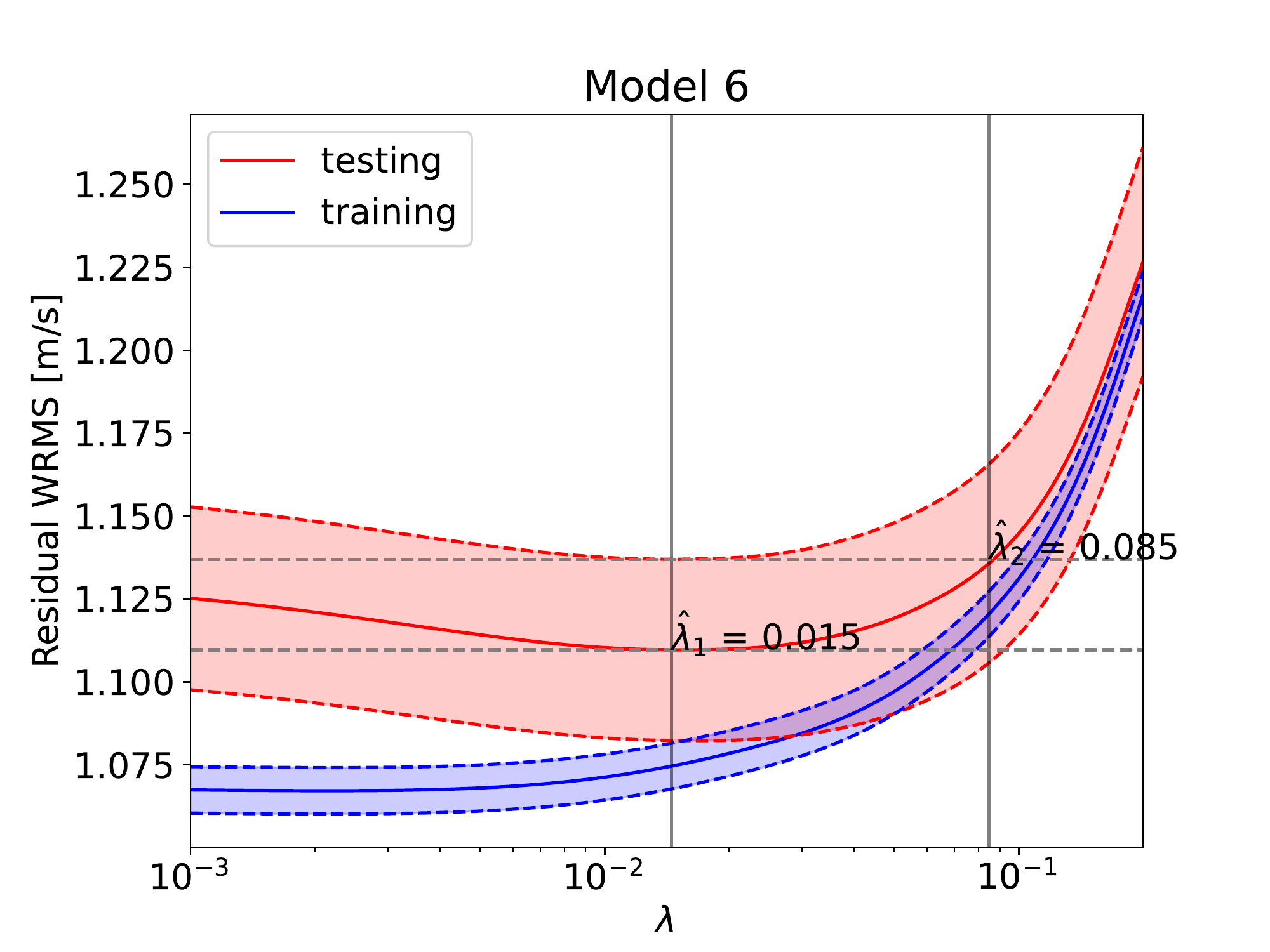}
    \includegraphics[width=0.49\columnwidth]{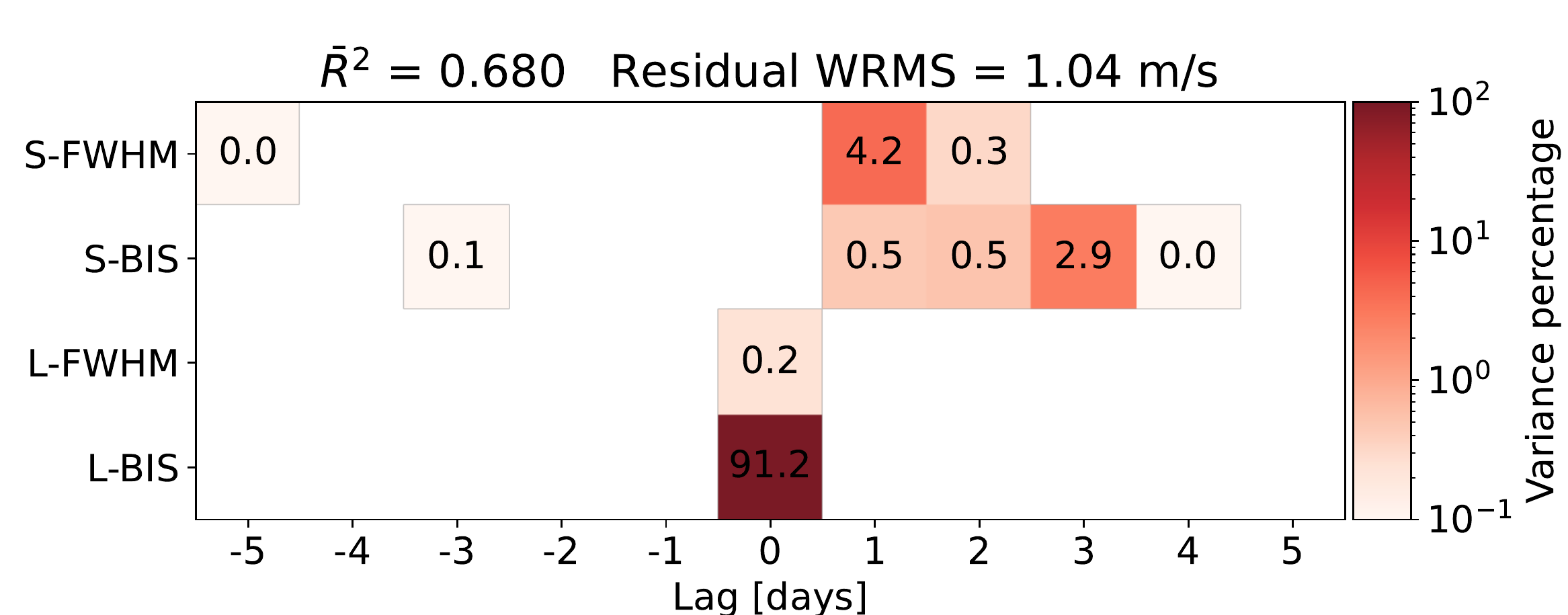}
    \caption{Same as Fig.~\ref{fig:lasso_var} but for FWHM and BIS. The regularised model with $\lambda=0.05$ returns $\bar{R}^2=0.680$ and a residual WRMS of 1.04~m/s.}
    \label{fig:lasso_fb}
\end{figure*}

Comparing Fig.~\ref{fig:ls-variation} and Fig.~\ref{fig:fwhm_bis}, $\textrm{L-PC}_1$ derived from \FIESTA appears to behave similarly to L-BIS, the long-term variation of the BIS timeseries. As Fig.~\ref{fig:lasso_fb} shows, this long-term components contribute to $\sim 91\%$ of variance in modelling $RV_\textrm{HARPS}$. The other $\sim9\%$ (equivalently 0.55~m/s RMS) comes from the short-term variations which we attribute to be primarily solar rotationally modulated variability. In terms of performance, using FWHM and BIS together results in a slightly worse $\bar{R}^2=0.680$ and a slightly larger residual weighed RMS of 1.04~m/s. This is likely due to the fact that the combination of FWHM and BIS is not as effective at describing the CCF as the combination of $\textrm{S-PC}_i$ and $\textrm{L-PC}_i$ derived from $\Delta RV_k$'s. 

%-------------
%-------------
% %-------------
% \begin{figure}[t]
%     \includegraphics[width=1\columnwidth]{figure/fwhm.png}
%     \includegraphics[width=1\columnwidth]{figure/BIS_SPAN.png}
%     \includegraphics[width=1\columnwidth]{figure/fwhm_Periodogram.png}
%     \includegraphics[width=1\columnwidth]{figure/bis_Periodogram.png}
%     \caption{Top: the unbinned and daily binned FWHM and BIS span timeseries of the three years of HARPS-N high-resolution spectroscopic data for the Sun. Bottom: their periodograms with prominent periods labelled in orange.}
% \label{fig:harps-N-fwhm_bis_daily}
% \end{figure}
% %-------------

\begin{figure*}[t]
\centering
    \includegraphics[width=0.55\columnwidth]{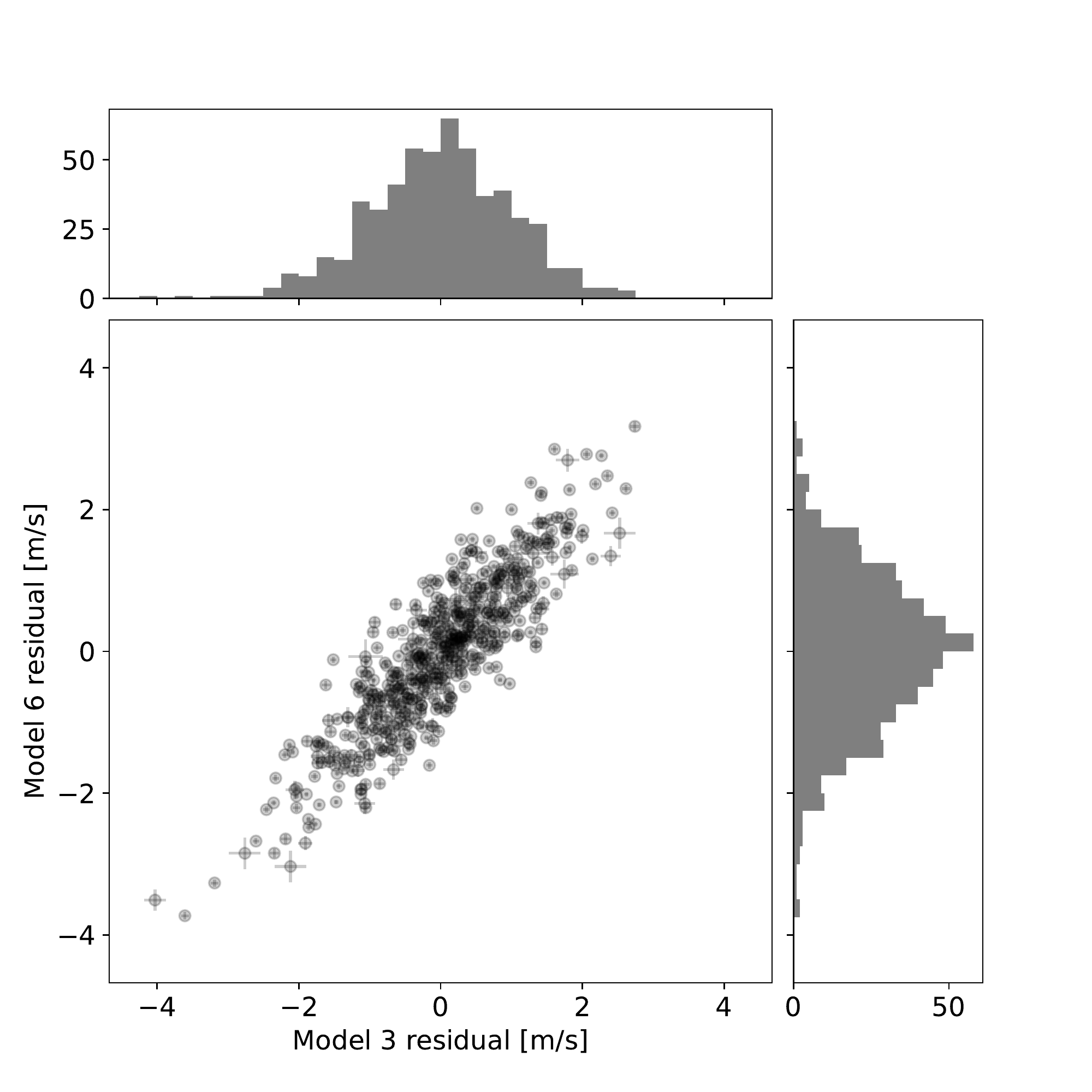}
    \includegraphics[width=0.44\columnwidth]{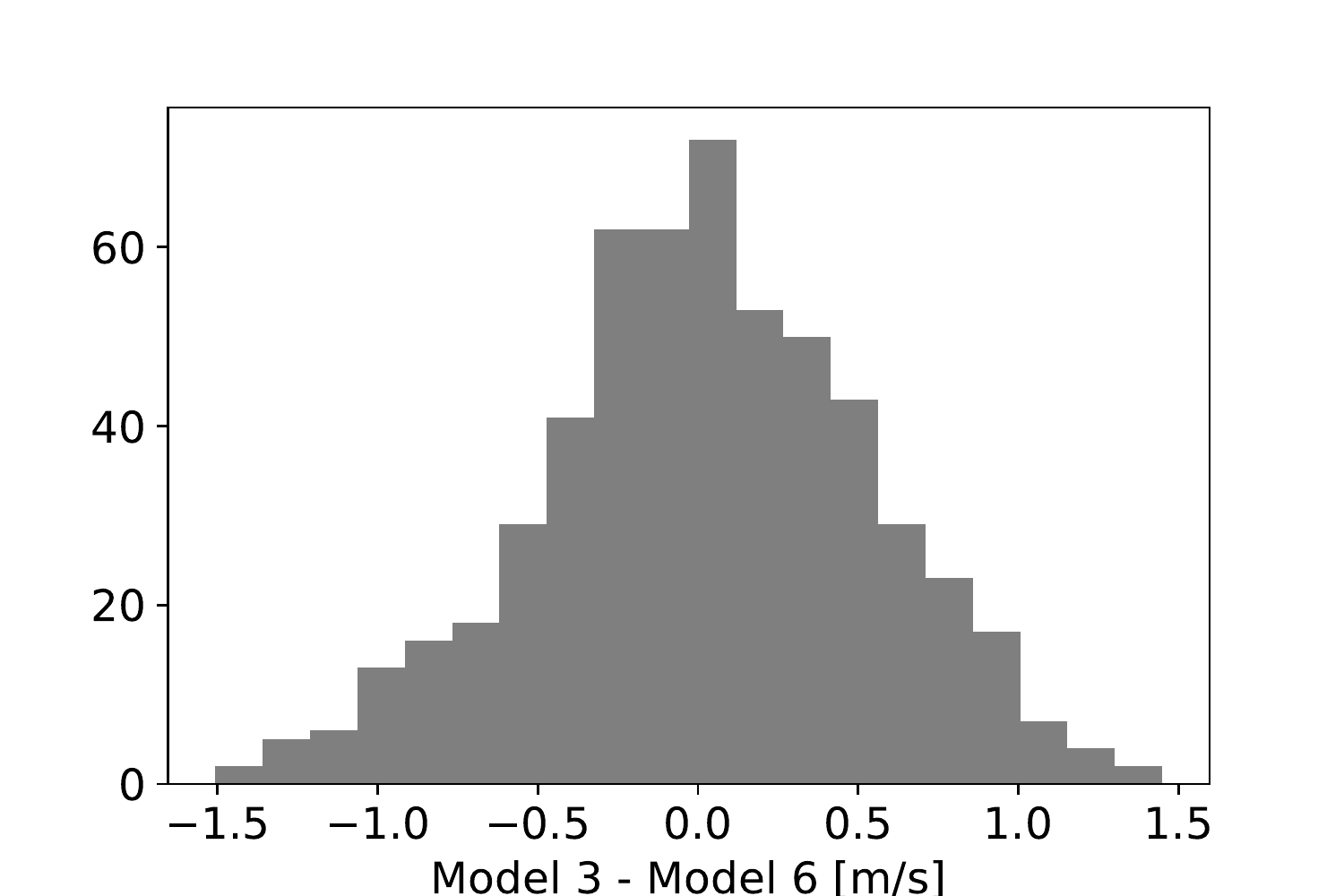}    
    \caption{Left: the correlation and histograms of residuals (i.e. the difference between the observations and the model predictions) using the Model 3 and Model 6 predictions show that the two models are consistent. Right: the difference between Model 3 and Model 6 predictions has a $1\sigma$ width of 0.77~m/s in the distribution.}
    \label{fig:residual_correlations}
\end{figure*}

%---------------------------------------------------
% \subsection{Final comparisons}
%---------------------------------------------------
We summarise 3 models using FWHM and BIS for predicting the HARPS-N solar spurious RV in Table~\ref{table:mlr1} (Models 4-6). In order to compare how the models are similar to each other, we compare the residual correlations between Model 3 and Model 6 specifically. Their residuals are well correlated with each other, indicating that the two models agree well on each other in predicting $RV_\textrm{HARPS}$. The characteristic $1\sigma$ width of the residuals (the half-width of the central 68.2\% credible interval) is 0.77~m/s (Fig.~\ref{fig:residual_correlations}), indicating the overall difference of the two model predictions.
We suggest using this approach for evaluating the consistency between two or more models in the planet detection or stellar variability modelling, such as model comparisons in the EXPRES Stellar-Signals Project \citep{LilyZhao2022}.
%-------------
%-------------

%---------------------------------------------------
\section{Summary and discussions}
\label{sec:discussions}
%---------------------------------------------------
\subsection{Overview}

We presented the improved Fourier phase spectrum analysis (FIESTA or \fiesta) for disentangling intrinsic RV shifts from apparent RV shifts caused by stellar variability and instrumental instability. 
\FIESTA provides summary statistics for each spectra in a timeseries based on projecting the CCF onto a truncated version of the Fourier basis functions.  This approach is motivated by the shift-invariant properties of the Fourier basis:
\begin{enumerate}
    \item A shift in the signal (a spectral line profile or a CCF in the context of radial velocity exoplanet detection) does not change the amplitudes of each Fourier basis function:
    \begin{equation}
        A_k(t_i) = A_k(t_0) \qquad \textrm{for each $k$}
    \end{equation}
    \item A true Doppler shift results in the same shift for each of the basis functions:
    \begin{equation}
        RV_{\text{FT},k=1} = RV_{\text{FT},k=2} = \dots = RV_\textrm{shift}
    \end{equation}
    and thus 
    \begin{equation}
        \Delta RV_{k=1} = \Delta RV_{k=2} = 
        \dots = 0.
    \end{equation}    
\end{enumerate}
As a result, we can use changes in the amplitude  $A_k$ and RVs measured as $RV_{\text{FT},k}$ or $\Delta RV_k$ for each Fourier mode $k$ to parametrise the variations of spectral line shapes. 
We focus on the analysis of $\Delta RV_k$ because it is directly related to the spurious RVs due to stellar and/or  instrumental variability.

Specifically, \FIESTA decomposes the spectral line profile CCF into the Fourier modes (Eq.~\ref{eq:CCF(v)_freq}), i.e. the Fourier basis functions truncated to have finite support (Fig.~\ref{fig:Flux_residual}, left panel). 
The amplitude ($A_k$) and the phase-derived RV shift ($RV_{\text{FT},k}$) for each Fourier mode $k$ fully parametrise the CCF through discrete Fourier transform. 
The suite of $RV_{\text{FT},k}$ is a $N$-dimensional measurement of the shared intrinsic planetary RV shift of the CCF plus any stellar or instrumental variability measured at mode $k$ (Section~\ref{sec:weighted_mean}). 
The changes with time of the amplitudes and RV shifts for each Fourier mode (denoted as $A_k$ and $\Delta RV_k$) is used to trace the CCF variability.

As discussed in Section~\ref{sec:Measurement_uncertainties}, higher Fourier modes generally come with larger uncertainties. Practical consideration to pick the number of modes for the analysis include (1) the precision of reconstructing the CCF, (2) the uncertainties of $A_k$ and $RV_{\text{FT},k}$ for a single observation, (3) the SNR of $A_k$ timeseries and $\Delta RV_{\text{FT},k}$ timeseries and (4) the instrument resolution. 

We demonstrated the use of \FIESTA in analysing the RV timeseries from Section~\ref{sec:FIESTA_SOAP} to Section~\ref{sec:BIS-FWHM}, covering SOAP~2.0 solar simulations and the 3-years of HARPS-N solar observations.

\paragraph{SOAP~2.0 simulations}
We demonstrated how \FIESTA distinguishes an intrinsic line shift from an apparent line shift due to a line deformation (Section~\ref{sec:challenging_test}). We also explored the response of $\Delta RV_k$ to simulated solar spots and plages at different latitudes, as well as their correlations with the apparent RV (Section~\ref{sec:spot_plage_correlations}). 
The simulated solar observations show high correlations between the spot-induced spurious RVs and the lower Fourier modes (Section~\ref{sec:Simulated continuous solar observations}).  

\paragraph{HARPS-N solar data}
In Section~\ref{sec:FIESTA_HARPS}, we applied \FIESTA to 3 years of HARPS-N solar observations (Fig.~\ref{fig:time-series_and_shift_correlation} and \ref{fig:time-series_and_A_correlation}).
As neighbouring Fourier modes may show high correlations with each other, we are motivated to use PCA for dimension reduction.
With PCA, we extracted the most prominent 3 orthogonal features from the first 9 $\Delta RV_k$ timeseries (Fig.~\ref{fig:FIESTA_RV_PCA}).
Next, we separated the short-term variability dominated by solar rotation from the long-term variability modelled by a Mat\'ern 5/2 GP kernel with a length scale of 100 days (Fig.~\ref{fig:ls-variation}). 
Feeding these features into multiple linear regression models  to fit $RV_\textrm{HARPS}$, we reduced the WRMS RV from 2.0~m/s to 1.14~m/s ($\bar{R}^2=0.67$) in a model that uses only the spectra at each epoch.
Incorporating \FIESTA outputs a few days prior/after each observations allowed us to further reduce the WRMS RV from 1.89~m/s to 0.94~m/s ($\bar{R}^2=0.73$). To avoid overfitting, Lasso regression and cross validation have been implemented.
Our model shows the solar rotationally induced variability contributes only 7\% to the total RV variation modeled by \FIESTA (Fig.~\ref{fig:lasso_var}).
The long-term variability, which we identified as sources from the solar magnetic cycle and instrumental instability, compose of over 80\% of the RV variations in $RV_\textrm{HARPS}$.
The remaining less than 10\% RV variability come from the CCF changing resulted from the apparent solar rotation rate changes due to Jupiter and Earth's eccentricity. 
Therefore, it is important for future EPRV exoplanet surveys to consider modelling and correcting for instrumental instability and long-term activity variations, in addition to modelling the rotationally-linked stellar variability.

Furthermore, we compared the models using PC scores of $\Delta RV_k$ as features (Section~\ref{sec:FIESTA_HARPS}) and the ones using FWHM and BIS (Section~\ref{sec:BIS-FWHM}) as summarised in Table~\ref{table:mlr1}. 
The latter performs slightly worse with a residual WRMS of 1.18~m/s ($\bar{R}^2=0.65$) without lags and 1.04~m/s ($\bar{R}^2=0.68$) with lags. The fact that FWHM and BIS do not perform as well may be because they do not provide as complete CCF parametrisation as the \fiesta-derived $\Delta RV_k$. 
However, both methods predict the solar rotational spurious RMS RV $\sim0.5$~m/s. As a measurement of model consistency, the RMS of the difference between the two residuals using these two models is $\sim0.77$~m/s (Fig.~\ref{fig:residual_correlations}). 
We encourage future studies to perform a similar consistency check for multiple methods proposed to reduce the effects of stellar variability.

%---------------------------------------------------
\subsection{Comparison to previous works}
%---------------------------------------------------

\subsubsection{Original \FIESTAtitle}

The original \FIESTA proposed by \cite{jzhao2020} extracted the CCF variability into the lower and higher frequency ranges $RV_\text{FT,L}$ and $RV_\text{FT,H}$. 
The current version of \FIESTA presented here decomposes the CCF into all the available Fourier modes up to the CCF sampling limit, providing a comprehensive parametrisation of the CCF using the amplitudes $A_k$ and the phase derived RV shift $RV_{\text{FT},k}$ and the CCF variability using $A_k$ and $\Delta RV_k$.
This facilitates the study of multiple sources contributing to CCF variability, as demonstrated in our analysis of the HARPS-N solar observations in Section~\ref{sec:FIESTA_HARPS}. 

%---------------------------------------------------
\subsubsection{\Scalpels}
%---------------------------------------------------

\cite{Cameron_2021} introduced \Scalpels to study the CCF variability using the autocorrelation function (ACF) of the cross-correlation function of each spectrum with a mask. 
\Scalpels was able to reduce the RMS of daily averaged HARPS-N solar spectra from 1.76~m/s to 1.25~m/s, a 29\% reduction in the RV scatter.
In order to make a more direct comparison to this result, We conducted a preliminary analysis of the same HARPS-N solar observations as \cite{Cameron_2021}. When applied to the 5 year timeseries, \FIESTA reduced the weighted daily RV RMS to 1.09~m/s using the same approach as in Section~\ref{sec:mlr_with_lags}.
Since the \citet{Cameron_2021} dataset provides only daily binned CCFs and does not include FWHM and BIS measurements and it became available after much of our analysis had been completed, this manuscript only focuses on the three year timeseries available on \url{https://dace.unige.ch/sun/} at the time writing. 

In addition to comparing the amount of stellar variability removed, it is interesting to compare the basis vectors used by the two studies. The eigenvectors for the ACF in \citet{Cameron_2021} Fig 3 look very similar to the Fourier basis functions used in \FIESTA (Fig.~\ref{fig:Flux_residual}).  
%though the ACF eigenvectors seem to flatten out towards the CCF edges, but this could be because a CCF shift at continuum has smaller derivative than at the CCF core and thus the edges providing less constraint on the ACF eigenvectors. 
It is exciting to see how the two completely different methods - the data driven approach from \cite{Cameron_2021} and the theoretical derivation from \cite{jzhao2020} and this paper end up with similar basis functions (i.e. eigenvectors). 
% As a result, their scores and our amplitudes \jzhao{which figures... ACF vs CCF} (both being coefficients projecting the solar CCFs onto the orthogonal basis) look similar in the timeseries. 

%---------------------------------------------------
\subsubsection{Machine learning}
%---------------------------------------------------
\cite{debeurs2020identifying} analysed 3 years HARPS-N solar observations and predict the spurious RV due to stellar variability by applying either linear regression or a dense neural network to the CCF.  Using a neural network, they reduced the RV scatter by $47\%$, very similar to our $48\%$ reduction in the WRMS using the multiple linear regression with \FIESTA indicators and allowing for a lag (Section~\ref{sec:mlr_with_lags}). A precise comparison of these two results is not practical because the \cite{debeurs2020identifying} HARPS-N RVs were based on an earlier version of the HARPS-N data reduction system, while ours are from the newer version (see discussions in Section~\ref{sec:data}). Regardless, there is room for further research in reducing the residual scatter by treating the \FIESTA inputs (either directly or after dimensional reduction via PCA)  
%approach with the advantage of separating various sources contributing spurious RVs and 
with neural network methods that are able to learn non-linear relationships between activity indicator proxies and RVs.  
%with the advantage of the flexibility of building up more layers for the model and without the assumption of linearity between the spurious RVs and the 

\subsection{Opportunities for Future Research}
%---------------------------------------------------
\subsubsection{PCA}
%---------------------------------------------------
We explored PCA for dimension reduction and used the PC scores as features to characterise the CCF deformation. 
As PCA is an unsupervised learning technique, we could not separate out the different mechanisms that caused the CCF deformation.
In order to improve interpretability, we computed a smoothed version of the PC scores ($\textrm{L-PC}_i$'s) that was insensitive to changes on the solar rotation timescale.
Indeed, this resulted in the long-term solar magnetic cycle and the CCD detector warm-up events becoming more prominent in $\textrm{L-PC}_1$.
Inspecting the difference between the PC scores and smoothed PC scores allowed us to separate out the short-term variability (e.g. the solar rotation). 
In future studies, we could explore employing regularisation in PC scores so as to drive some basis vectors to focus on variations on a timescale near the stellar rotation period and other basis vectors to focus on CCF shape changes occurring on longer timescales (e.g. instrumental variations or long-term stellar cycles).

%---------------------------------------------------
\subsubsection{Gaussian processes}
\label{sec:gp}
%---------------------------------------------------
We used activity indicator proxies (PC scores derived from $\Delta RV_k$; FWHM and BIS) as regressors in an effort to model the spurious RV introduced by solar variability.
Although they depend only on each instantaneous spectrum, we showed that incorporating temporal information by performing regression on these activity indicator proxies and the lagged components simultaneously resulted in improved prediction power, as evidenced by $\bar{R}^2$. 
This implies Gaussian process (GP) regression, particularly, multivariate GP models with a common latent process and its derivatives approximating the effect of regressing on the lagged indicator proxies would be a promising approach. (\citealp{Rajpaul_2015, jones2020improving}; \citealp[GLOM,][]{Gilbertson_2020_II}. 
Although GP may not be needed for the highly sampled solar data analysed in this manuscript, planet hunting in other stellar targets with sparse observing cadence would benefit from GP analysis of both the stellar spurious RVs and the activity indicator proxies \citep[e.g.][]{Langellier2021, Faria2022}.
In fact, we have made initial attempts to combine the PC scores derived from $\Delta RV_k$ and the multi-variate GP known as GLOM for modelling the stellar variability for HD~101501, HD~34411, HD~217014, and HD~10700 as part of the EXPRES Stellar-Signals Project \citep{LilyZhao2020}.
This approach allows \FIESTA to be applied to stars other than the Sun, for which gaps in the observations are much more common \citep{LilyZhao2022}. 

Most published applications of GP regression to radial velocity timeseries have assumed stational GP kernels.  Stationary kernels are not well suited for modelling sudden changes in the $RV_\textrm{HARPS}$ due to detector warm-ups or power failures.  
We find that these contribute a significant fraction of the CCF deformation changes in the HARPS-N solar dataset.    
Future studies may benefit from adopting more complex GP kernels, such as the sum of a stationary GP kernel for modelling stellar variability and a non-stationary kernel with parameters informed by prior knowledge about any instrumental changes.  

%---------------------------------------------------
\subsubsection{CCF construction}
%---------------------------------------------------
In this paper, we demonstrate the \FIESTA methodology using HARPS-N solar observations and simulated SOAP data. 
Publicly available CCFs for HARPS-N solar data and SOAP data are based on a weighted binary mask. 
While CCFs based on a template spectrum can improve the robustness of RVs for some stars (e.g., low SNR, broad absorption features in cool stars), we expect a narrow, weighted binary mask to preserve more line shape information.
While it could be interesting to study the effects of different CCFs masks and preprocessing (e.g., CCF continuum normalisation, velocity range) on \FIESTA outputs, that is beyond the scope of this paper. Of course, one should use the CCFs constructed using the same mask and preprocessing to correctly account for other changes due to stellar variability and instrumental instability.

%---------------------------------------------------
\subsubsection{Improving \texorpdfstring{$\Delta RV_k$}{Lg}}
%---------------------------------------------------
% EBF: This section would need to be completely rewritten.

$\Delta RV_k$, the difference between the RV shift at $k$-th Fourier mode ($RV_{\text{FT},k}$) and the apparent RV ($RV_{\mathrm{apparent}}$), traces the amount of CCF variability at the frequency $\xi_k$ (Eq.~\ref{eq:delta_RV_k}). 
By construction, a true Doppler shift ($RV_{\text{planets}}$) does not affect the expected values of individual $\Delta RV_k$'s, but contributes to each of the $RV_{\text{FT},k}$.
%
%However, part of the RV due to intrinsic line variability is also removed in the subtraction as it is present in both of these two terms. 
%
For the HARPS-N solar spectra, there are no true Doppler shift signals in the heliocentric RV timeseries and so $RV_{\text{FT},k}$ alone would trace the CCF variability and could have been a stronger indicator for variability. 
Nevertheless, for other stars, there will inevitably be uncertainty about the potential presence of planetary companions.
That is why we choose to use $\Delta RV_k$ instead of $RV_{\text{FT},k}$ even for the analysis for the HARPS-N solar observations as a manner of consistency.

It may be possible to improve the statistical power and accuracy of $\Delta RV_k$ as a series of CCF variability indicators by re-defining 
\begin{equation}
    \widehat{\Delta RV_k}^{(l)} \equiv RV_{\text{FT}, k} - RV_{\text{planets}}^{(l)} \qquad k=0, 1, \dots, N-1
\label{eq:delta_RV_k2}
\end{equation}
once we have a first guess of $RV_{\text{planets}}^{(1)}$, the RV signal of putative planetary companion(s).
Such first guess may be derived from jointly fitting the planetary RV signal and intrinsic stellar variability with $\Delta RV_k$.
Then as an iterative approach, we keep improving the fit of $RV_{\text{planets}}^{(l+1)}$ as we improve $\widehat{\Delta RV_k}^{(l)}$ from the previous round. 

%by using a planetary model for the RV affecting each $RV_{k}$ as a function of time, rather than estimating the true RV from each observation individually.  

%One approach to analyzing a spectroscopic timeseries of a potential exoplanet host star is to iteratively apply \FIESTA to the residuals between the apparent RVs and the current planetary model.  
%With the iterative approach, we can jointly fit the planetary RV signal and intrinsic stellar variability.
% Initially, we compute $\widehat{\Delta RV_k}^{(0)}$ as described in \S~X.X.  
% At each iteration, we can fit for the RV signal of an putative planetary companions, $RV_{\text{planets}}^{(l+1)}$ using the residuals of 
% $RV_{\mathrm{apparent}} - \widehat{\Delta RV_k}^{(l)}$. 
% Then we update the \FIESTA indicators $\Delta RV_k^{(l+1)}$
%as the CCF variability tracer by re-defining 

%RV residuals to the  re-defined $\widehat{\Delta RV_k}$ can then be used to jointly fit the planets and stellar variability. 
% The process above can be iterated until $\widehat{\Delta RV_k}^{(l)}$ and $RV_{\text{planets}}^{(l)}$ converge.

For its implementation, one could take a Bayesian approach and simultaneously analyse the planetary signal and stellar variability.
This would require placing priors on both the planet properties and the \FIESTA indicators (e.g. $A_k$'s and $\Delta RV_k$'s).  
Multivariate Gaussian (GP) regression (Section~\ref{sec:gp}) provides a powerful framework for placing priors on the \FIESTA indicators.
It is beyond the scope of this paper and offers opportunities for future research.

\section{acknowledgments}
The authors gratefully acknowledge Suvrath Mahadevan, Jessi Cisewski Kehe, Nikunj Sura and the Penn State team for their direct and indirect input to the methodology, coding and writing of this manuscript.
The analysis of this work used the public three years of HARPS-N solar data hosted by University of Geneva.
We thank Xavier Dumusque and Zoe de Beurs for their opinions on the 3 years HARPS-N solar observations.
We also thank Andrew Collier Cameron for the productive discussions on \Scalpels and the 5 years HARPS-N solar observations. 
J.Z. is grateful to Petra Br\v{c}i\'{c} for the inspirational discussions on the mathematical part of the \FIESTA method over the course of preparing the paper.
J.Z. thanks Ludovic Delchambre and Jake Vanderplas for clarifying the weighted PCA theories and implementations.

This research was supported by Heising-Simons Foundation Grant \#2019-1177.  
E.B.F. acknowledges the support of the Ambrose Monell Foundation and the Institute for Advanced Study.  
This work was supported by a grant from the Simons Foundation/SFARI (675601, E.B.F.).
We acknowledge the Penn State Center for Exoplanets and Habitable Worlds, which is supported by the Pennsylvania State University and the Eberly College of Science. 
% Computations for this research were performed on the Pennsylvania State University's Institute for Computational and Data Sciences' Roar supercomputer."

We acknowledge the use of the following software for producing the results of this research:
\software{
\texttt{astropy} \citep{Astropy2013, astropy2018}, 
SOAP~2.0 \citep{Dumusque2014SOAP}, 
\texttt{wpca} (\url{https://github.com/jakevdp/wpca}), 
\texttt{george} \citep{Ambikasaran2015}.
}

The Github repository for the analysis linked in the paper including the \FIESTA code is \url{https://github.com/jinglinzhao/FIESTA-II}, which is also available in Zenodo \citep{FIESTA_code}.

% \ebf{Can cite deBuers paper that found linear regression on the full CCF did nearly as well as their neural network.  Would you want to include linear regession on the full CCF as a point of comparison?  (The de Buers paper got a slightly lower RV RMS, partialy because they used an older reduction of HARPS-N data that had errorneously subtracted out a long-term trend in RVs.  So comparing your results to the value from their abstract won't be entirely apples-to-apples.}

% \jzhao{prediction RV vs the observed RV - good or bad model}

% \jzhao{write median errorbar for each k}

% \jzhao{list in a table the breakdown frequency for SOAP and HARPS-N}

\newpage
\appendix


%%%%%%%%%%%%%%%%%%%%%%%%%%%%%%%%%%%%%%%%%%%%%%%%%%%%%
\section{The coefficient of determination $R^2$ and the adjusted $R^2$}
\label{sec:coefficient_of_determination}
%%%%%%%%%%%%%%%%%%%%%%%%%%%%%%%%%%%%%%%%%%%%%%%%%%%%%
The coefficient of determination, also known as $R^2$, is defined as $1-\frac{\textrm{wRSS}}{\textrm{wTSS}}$ where wRSS is the weighted sum of squares of residuals and wTSS is the total weighted sum of squares \citep{Heinisch1962SteelRG}. It is a measurement of the percentage of variance of the dependent variable predicted by the independent variable(s), with $R^2=1$ indicating a zero residual and the modelled values totally match the observations. 

To avoid overestimating when extra explanatory variables are introduced into the model, the adjusted $R^2$, denoted as $\bar{R}^2$, is used \citep{Raju_1997}. $\bar{R}^2$ is defined as $1-(1-R^2)\frac{n-1}{n-p-1}$ where $n$ is the sample size and $p$ is the number of explanatory variables. $\bar{R}^2$ can thus be used to compare the performance of different models, regardless of different explanatory variables being used. For the simple linear regression, there is only a single explanatory variable and so $p=1$.

%%%%%%%%%%%%%%%%%%%%%%%%%%%%%%%%%%%%%%%%%%%%%%%%%%%%%
\section{Noise in amplitude and phases}
\label{sec:noise}
%%%%%%%%%%%%%%%%%%%%%%%%%%%%%%%%%%%%%%%%%%%%%%%%%%%%%
For simplicity, we treat photon noise across the velocity grid as uncorrelated and Gaussian (approximated for Poisson distribution because of high photon counts), with the mean being the observed flux $CCF(v_n)$ and the $1\sigma$ of the Gaussian distribution being the associated photon noise error counts at each $v_i$ denote as $\sigma_{CCF(v_n)}$. 
Then the uncertainty of $\widehat{CCF}(\xi_k)$ can also be calculated using DFT
\begin{equation}
    \sigma_{\widehat{CCF}(\xi_k)} = \sum_{n=0}^{N-1} \sigma_{CCF(v_n)} e^{-i\frac{2\pi}{N}nk} \qquad k=0, 1, \dots, N-1.
\label{eq:d_DFT}
\end{equation}
Since the DFT is a linear transform, a normally distributed $CCF(v_n)$ results in a normally distributed $\widehat{CCF}(\xi_k)$. 
The distribution of $\widehat{CCF}(\xi_k)$ can be approximated by a 2D multivariate normal distribution in the Fourier domain. The marginalised $\widehat{CCF}(\xi_k)_\text{RE}$ and $\widehat{CCF}(\xi_k)_\text{IM}$ also follow the normal distribution, with the variance
\begin{equation}
    \sigma^2_{\widehat{CCF}(\xi_k)_\text{RE}} \approx \sigma^2_{\widehat{CCF}(\xi_k)_\text{IM}} \approx \frac{1}{2} \sum_{n=0}^{N-1} \sigma^2_{CCF(v_n)}.
\end{equation}

\begin{figure*}[t!]
\centering
    \includegraphics[width=0.16\columnwidth]{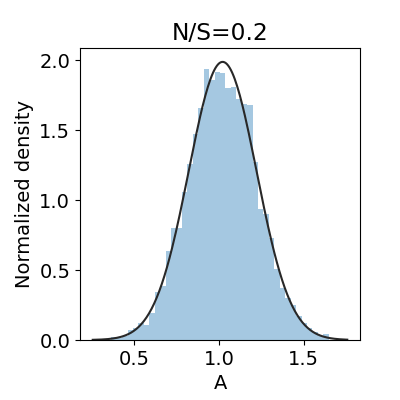}
    \includegraphics[width=0.16\columnwidth]{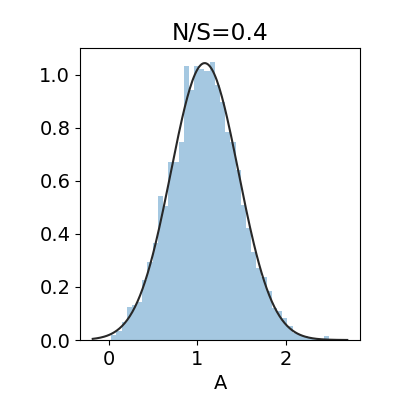}    
    \includegraphics[width=0.16\columnwidth]{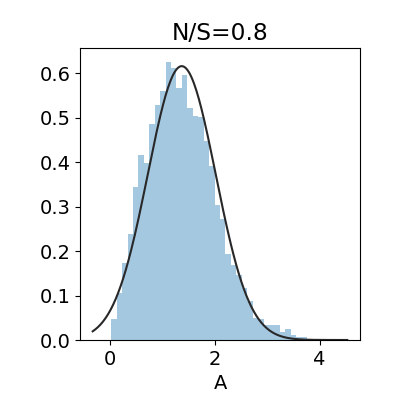}
    \includegraphics[width=0.16\columnwidth]{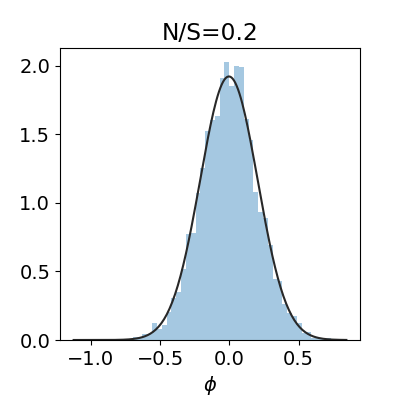}
    \includegraphics[width=0.16\columnwidth]{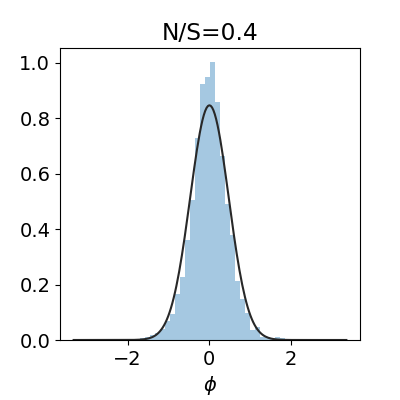}    
    \includegraphics[width=0.16\columnwidth]{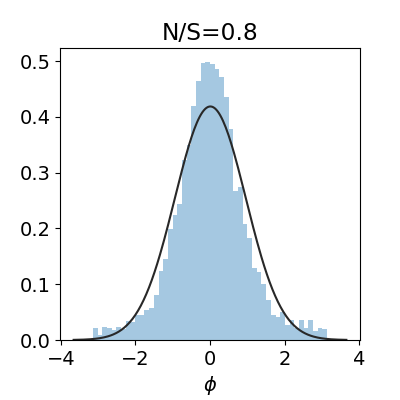}
    \caption{The distributions of the amplitudes and phases in the presence of noise, where the signal (S) refers to $|\widehat{CCF}(\xi_k)| = 1$ and the noise (N) refers to $\sigma_{\widehat{CCF}(\xi_k)_\text{RE}}$ and $\sigma_{\widehat{CCF}(\xi_k)_\text{IM}}$. The black curve is a normal distribution fit to the histogram (regardless whether the underlying distribution is normal). N/S~=~0.2 is the threshold when the distributions of $A_k$ and $\phi_k$ start to deviate from normal distributions under the normality test. For comparison, examples of N/S~=~0.4 and 0.8 represent the lower SNR regime, where the $A_k$ distributions tend to be skewed and the $\phi_k$ distributions tend to be taller in the core and thinner in the wing than a normal distribution.}
    \label{fig:normality_test}
\end{figure*}

Next we consider what distributions the amplitudes $A_k$ and $\phi_k$ as calculated in Eq.~\ref{eq:A_k} and \ref{eq:phi_k} follow.
We study under what condition $A_k$ and $\phi_k$ are well-approximated by the normal distribution. Without loss of generality, we choose $\widehat{CCF}(\xi_k) = (1, 0)$ and test in a bootstrapping approach if the distribution of $A_k$ and $\phi_k$ can be distinguished from a normal distribution when the noise-to-signal-ratio $N/S=\sigma_{\widehat{CCF}(\xi_k)_\text{RE}} = \sigma_{\widehat{CCF}(\xi_k)_\text{IM}} = 0.1,~0.2,~,\dots,~1$ using the D'Agostino`s K-squared \citep{Ralph1990} and the Shapiro-Wilk normality tests \citep{SHAPIRO}. 
We find that when $N/S \le 0.2$, the resulting $A_k$ and $\phi_k$ distributions are indistinguishable from a normal distribution. Larger $N/S$ results in deviations from the normal distribution (Fig.~\ref{fig:normality_test}).

%-------------
%-------------

In conclusion, we compute the uncertainties of $\sigma_{\widehat{CCF}(\xi_k)_\text{RE}}$ and $\sigma_{\widehat{CCF}(\xi_k)_\text{IM}}$, which can be approximated by $\sqrt{\frac{1}{2} \sum_{n=0}^{N-1} \sigma^2_{CCF(v_n)}}$, and compare it with $|\widehat{CCF}(\xi_k)|$. When the ratio is less than 0.2, we can be confident that the distribution for the amplitudes and phases is well described by a normal distribution.

Since we treat noise across CCF pixels uncorrelated, the resulting uncertainties in the amplitudes and phases obtained by the bootstrapping are likely to be overestimated compared to a noise model that accounts for correlations in the CCF.

\section{Zero-padding}
\label{appendix:zero-padding}
%%%%%%%%%%%%%%%%%%%%%%%%%%%%%%%%%%%%%%%%%%%%%%%%%%%%%
For a sequence of $N$ numbers $\{x_n\}$ in the time domain and $\{X_k\}$ in the Fourier domain, the DFT simply follows
\begin{equation}
    X_k = \sum_{n=0}^{N-1}x_n \exp({-\frac{2\pi i}{N} kn}).
\label{eq:DDT}
\end{equation}

Zero-padding adds zeros to the end of the signal. Let $N$ be the intrinsic number of inputs of the original discrete signal and $N'$ be the total number of inputs for the DFT with $(N'-N)$ zeros padded after the original signal. Denote the new signal as $\{x'_n\}$.  According to Eq.~\ref{eq:DDT}, the Fourier transform of the zero-padded signal is 
\begin{equation}
\begin{split}	
    X'_k 
    &= \sum_{n=0}^{N'-1}x'_n \exp({-\frac{2\pi i}{N'} kn}) \\
    &= \sum_{n=0}^{N-1}x_n \exp({-\frac{2\pi i}{N'} kn}) + 
    \sum_{n=N}^{N'-1} 0 \exp({-\frac{2\pi i}{N'} kn}) \\
    &= \sum_{n=0}^{N-1}x_n \exp({-\frac{2\pi i}{N'} kn})
\end{split}	
\label{eq:DDT2}
\end{equation}
Replace $k/N$ in Eq.~\ref{eq:DDT} by $\xi$ and and $k/N'$ Eq.~\ref{eq:DDT2} by $\xi'$, we have 
\begin{equation}
    X_k = \sum_{n=0}^{N-1}x_n \exp({-2\pi i\xi n})
\end{equation}
and 
\begin{equation}
    X'_k = \sum_{n=0}^{N-1}x_n \exp({-2\pi i\xi' n}).
\end{equation}
For $N'>N$, $\xi'$ means a finer sampling than $\xi$ in the frequency grid. Therefore, the resulting amplitudes and phases in the Fourier transform space look smoother, but it does not extract more features of the original signal.

% \begin{figure}[!h]
% \centering
%      \begin{subfigure}[b]{1\textwidth}
%         \includegraphics[width=1\columnwidth]{figure/zero-padding-gaussian.png}
%     \end{subfigure}
%     \hfill
%      \begin{subfigure}[b]{1\textwidth}
%         \includegraphics[width=1\columnwidth]{figure/zero-padding-sin.png}
%     \end{subfigure}    
%     \caption{``Zero-padding a signal does not reveal more information about the spectrum, but it only interpolates between the frequency bins that would occur when no zero-padding is applied.'' [\url{https://dspillustrations.com/pages/posts/misc/spectral-leakage-zero-padding-and-frequency-resolution.html}] It helps reveal the actual frequency composition because the peak of the power spectrum is more accurately determined. The vertical line in the lower right panel is the frequency of the input signal.}
% \end{figure}
% %-------------


\bibliography{sample631}{}
\bibliographystyle{aasjournal}


\end{CJK*}
\end{document}